\newcommand{\eqnref}[1]{Eq.\ (\ref{#1})}
\newcommand{\eqnrefs}[2]{Eqs.\ (\ref{#1})--(\ref{#2})}
\newcommand{\eqnarefs}[2]{Eqs.\ (\ref{#1}) and~(\ref{#2})}
\newcommand{\secref}[1]{Sec.\ \ref{#1}}
\newcommand{\secsref}[2]{Secs.\ \ref{#1} and~\ref{#2}}

\newcommand{\figref}[1]{Fig.\ \ref{#1}}
\newcommand{\figsref}[2]{Figs.\ \ref{#1} and~\ref{#2}}
\newcommand{\figdref}[2]{Figs.\ \ref{#1} -- \ref{#2}}
\newcommand{\subfigref}[2]{Fig.\ \ref{#1}(#2)}
\newcommand{\bigO}[1]{$O(#1)$}
\newcommand{\Ham}{\mathcal{H}}
\newcommand{\etal}{\emph{et al}.}
\newcommand{\ie}{\emph{i.e.}}
\newcommand{\myfig}[5]{
\begin{figure}[{#5}]
\includegraphics[keepaspectratio,width=#4,angle=0]{#1}%\par
\begin{center}\begin{minipage}[b]{3.4in}{\caption{#2} \label{#3}}\end{minipage}\end{center}%
\end{figure}}

\documentclass[aps,pre,twocolumn,showpacs,amsmath,amssymb]{revtex4}
\usepackage{graphicx,dcolumn,bm,verbatim,color,hyperref}

\begin{document}

\bibliographystyle{apsrev} % Choose Phys. Rev. style for bibliography

%\preprint{APS/123-QED}
\title{Multiresolution community detection for megascale networks\\
by information-based replica correlations}

\author{Peter Ronhovde} %\email{ronhovde@hbar.wustl.edu}
\author{Zohar Nussinov} %\email{zohar@wuphys.wustl.edu}
\affiliation{Washington University in St. Louis, %Department of Physics, 
Campus Box 1105, 1 Brookings Drive, St. Louis, Missouri 63130, USA}%

\date{\today}

\begin{abstract}
We use a Potts model community detection algorithm 
to accurately and quantitatively evaluate the hierarchical 
or multiresolution structure of a graph.
Our multiresolution algorithm calculates correlations 
among multiple copies (``replicas'') of the same graph over 
a range of resolutions. 
Significant multiresolution structures are identified 
by strongly correlated replicas.
The average normalized mutual information, 
the variation of information, and other measures in principle
give a quantitative estimate of the ``best'' resolutions 
and indicate the relative strength of the structures in the graph.
Because the method is based on information comparisons, 
it can in principle be used with any community detection 
model that can examine multiple resolutions.  
Our approach may be extended to other optimization problems.
As a local measure, our Potts model avoids 
the ``resolution limit'' that affects other popular models.
With this model, our community detection algorithm has an 
accuracy that ranks among the best of currently available 
methods.
Using it, we can examine graphs over $40$ million nodes
and more than one billion edges.
We further report that the multiresolution variant 
of our algorithm 
can solve systems of at least $200~\!000$ nodes and $10$ 
million edges on a single processor with exceptionally high accuracy. 
For typical cases, we find a super-linear scaling, 
\bigO{L^{1.3}} for community detection and \bigO{L^{1.3}\log N} 
for the multiresolution algorithm where $L$ is the number 
of edges and $N$ is the number of nodes in the system. 
\end{abstract}

% # 89.75.Hc is more tangential 
%PACS, the Physics and Astronomy Classification Scheme.
%\keywords{Suggested keywords}

%BEGIN BODY TEXT
\maketitle{}

\vskip 0.1in

\section{Introduction}
\label{sec:intro}

One focus in the study of complex networks
is identifying suspected internal structure, 
and one characterization of such structure is in terms 
of ``community'' divisions within a model graph.
A recent introduction to the ``physics of networks'' can be  
found in~\cite{ref:newmanphystoday}.
One feature of organized structure within these systems 
is that the community divisions can depend 
on the scale at which the system is examined. 
Different scales correspond to distinct community divisions 
at different internal community edge densities.
For many systems, including those with hierarchical organization, 
a ``multiresolution'' approach~\cite{ref:multiresusage} 
is needed to capture the overall structure and the relationships 
between the elements at different resolutions.
Examples of such systems can include 
biological processes~\cite{ref:ravaszscience,ref:salespardo}, 
food webs~\cite{ref:clausetmissinglinks}, 
air transportation networks~\cite{ref:salespardo},
and communication networks~\cite{ref:blondel}.
Thus, multiresolution methods are an important extension 
of problems in community detection.

Some measures and methods regarding community detection 
are reviewed in~\cite{ref:albert,ref:fortunatosummary}.
Quality functions include modularity defined by 
Newman and Girvan~\cite{ref:gn},
a Potts model originally proposed 
by Reichardt and Bornholdt (RB)~\cite{ref:reichardt,ref:smcd},
our Potts model~\cite{ref:rz} that eliminates 
the random partition applied by RB, 
an application of a Potts model utilizing a mean-field 
approximation with ``belief propagation''~\cite{ref:hastings}, 
and another measure ``fitness''~\cite{ref:lanc}.
Other approaches include clique 
percolation~\cite{ref:palla,ref:kumpulacliqueperc}, 
spectral~\cite{ref:newmanmatrix},
continuous mapping to a conic optimization 
problem~\cite{ref:hildebrand}, 
``label propagation''~\cite{ref:LPA,ref:barber}, 
dynamical~\cite{ref:gudkov,ref:boccaletti},
and maximum likelihood~\cite{ref:clausetML}.
Karrer \etal{} \cite{ref:karrer} defined a measure of robustness 
of community structure based on random perturbations.
Some efforts enhance or expand applications 
to more general systems such as weighted 
networks~\cite{ref:newmanweighted,ref:smcd,ref:rz},
heterogeneous systems~\cite{ref:danonhetero,ref:rz},
bipartite graphs~\cite{ref:guimerabipartite,ref:duwuwang},
overlapping 
nodes~\cite{ref:overlapdef,ref:palla,ref:lanc,ref:duwuwang,ref:reichardt},
and multiresolution methods.

The multiresolution algorithm presented in this paper 
({\bf{1}})~determines and \emph{quantitatively} evaluates 
the relative strength of multiresolution structure(s) 
within a graph by examining the correlations among several 
independent solutions (``replicas'') of the same graph over 
a range of resolutions.
Strong correlations in the normalized mutual information (NMI) 
or the variation of information (VI) indicate the ``best''
system resolutions, and the relative value of the measure 
gives a quantitative estimate of the strength of the structures.
This quantitative evaluation of the best resolution(s) 
for the system is lacking or missing in most other multiscale 
community detection algorithms.
({\bf{2}})~The method is not limited to hierarchical structures 
but applies to general structures at different scales.
({\bf{3}})~Our approach is based on relative information 
comparisons, so it can in principle be used with any community 
detection model that can target different resolutions.
({\bf{4}})~The underlying Potts model and community detection
algorithm demonstrate an accuracy at least equal to the best 
methods currently available 
(see Appendix \ref{app:cdaccuracy})~\cite{ref:rz}.
The model is robust to the effects of noise 
(see Appendices A and B), and as a local measure, 
it is free of the 
``resolution limit''~\cite{ref:resolutionlimitdef} 
as discussed in the 
literature~\cite{ref:gn,ref:fortunato,ref:smcd,ref:kumpula}. 
%({\bf{5}})~The multiresolution algorithm builds on the accuracy 
%of the community detection algorithm,
%and we demonstrate that is extremely accurate 
({\bf{5}})~With improvements discussed in \secref{sec:cdalgorithm}, 
%our community detection algorithm is competitive 
it is competitive 
with the best algorithms currently available both in terms 
of speed and possible system size.
A single community solution can achieve systems as large 
as $40$ million  nodes and one billion edges with 
a computational time of $3.7$ hours 
(see Appendix \ref{app:cdlarge})~\cite{ref:computerused}.
({\bf{6}})~Our multiresolution algorithm is extremely accurate 
for large systems (see \secref{sec:accuracy}).
%({\bf{7}})~Our multiresolution algorithm is limited more 
%by computational time. 
%We apply it to systems up to $200~\!000$ nodes and over 
%$10$ million edges with a run time of about $4.6$ 
%hours on a single processor~\cite{ref:computerused}.
({\bf{7}})~We apply it to \emph{megascale} systems 
with over $10$ million edges and $200~\!000$ nodes 
with a run time of about $4.6$ hours on a single 
processor~\cite{ref:computerused}.
The algorithm should adapt very efficiently to parallel  
or distributed computing methods enabling larger systems 
to be studied.

Hierarchical organization is the most obvious type 
of multiresolution structure.
Some earlier work on hierarchies in graphs can be found 
in~\cite{ref:ravaszbarabasi,ref:ravaszscience}.
Examples of more recent efforts in analyzing hierarchical 
structures in graphs 
are~\cite{ref:hierdescriptors,ref:blondel,ref:lanc,ref:salespardo,ref:multires}.
Arenas \etal{} \cite{ref:multires} defined a multiresolution
method using modularity that makes novel use of the resolution 
limit~\cite{ref:resolutionlimitdef}.
Reichardt and Bornholdt~\cite{ref:reichardt}, 
Arenas \etal{}~\cite{ref:multires}, 
Kumpula and co-workers~\cite{ref:kumpulamultires}, 
and Heimo \etal{}~\cite{ref:heimo} 
also study multiresolution applications of an RB Potts model.

In this paper we will show, for the first time, how information 
theory based measures may be used to systematically extract the 
best community partitions on all scales.
This will enable us to methodically determine the hierarchical
or multiresolution structure of arbitrary networks.
In \secref{sec:informationmeasures}, we first briefly 
review the information measures that we employ.
Then in \secsref{sec:model}{sec:cdalgorithm}, 
we briefly discuss our Potts model and community 
detection algorithm, followed by an explanation 
of their applications to multiresolution analysis 
in \secref{sec:algorithm}. 
We then present several examples in \secref{sec:examples}.
The exceptional accuracy of the multiresolution algorithm 
is addressed in \secref{sec:accuracy}, 
and we conclude in \secsref{sec:discussion}{sec:conclusion}.
Details concerning the high accuracy and large size limit 
of the underlying community detection algorithm are 
relegated to Appendixes \ref{app:cdaccuracy} and 
\ref{app:cdlarge} respectively.
Appendix \ref{app:cdtransition} demonstrates an example 
of new transition effects in community detection
(such transitions directly affect replica correlations).
Appendix \ref{app:generalentropy} explains a 
generalization of our replica method for other, 
nongraph theoretical, optimization problems.
Appendix \ref{app:lfkdetails} elaborates on some details 
related to the benchmark accuracy test discussed 
in \secref{sec:accuracy}.

\section{Information Theory Measures}
\label{sec:informationmeasures}

The normalized mutual information $I_N$ and the 
variation of information $V$ provide methods 
of comparing one proposed community division to another.
In order to define $I_N (A,B)$ or $V(A,B)$ between 
two partitions $A$ and $B$, 
we first ascribe a Shannon entropy $H(A)$
for an arbitrary community partition $A$.
We assign the probability that a given node will fall 
in community $k$ as $P(k) = n_k/N$, 
where $n_k$ is the number of nodes in community $k$
and $N$ is the total number of nodes in the system. 
Then the Shannon entropy is
\begin{equation} \label{eq:shannonentropy}
  H(A) = -\sum_{i=1}^{q_A} \frac{n_k}{N} \log\frac{n_k}{N}
\end{equation}
where $q_A$ is the number of communities in partition $A$.

Mutual information $I(A,B)$ 
was developed within information theory.
It evaluates how similar two data sets 
are in terms of information contained in both sets of data.
The mutual information between two partitions $A$ and $B$ 
of a graph is calculated by defining a ``confusion matrix'' 
for the two community partitions.
The confusion matrix specifies how many nodes $n_{ij}$ 
of community $i$ of partition $A$ are in community $j$ 
of partition $B$.  
Mutual information $I(A,B)$ is defined as 
\begin{equation} \label{eq:mi}
  I(A,B) = 
  \sum_{i=1}^{q_A}\sum_{j=1}^{q_B} \frac{n_{ij}}{N} 
  \log\left(\frac{n_{ij} N}{n_i n_j}\right)  
\end{equation}
%where $N$ is the total number of nodes in the system,
where $n_i$ is the number of nodes in community $i$ of partition $A$
and $n_j$ is the number of nodes in community $j$ of partition $B$.
%and $q_A$ and $q_B$ are the number of communities in 
%partitions $A$ and $B$.
An interesting generalized mutual information is also 
defined in~\cite{ref:generalentropy}.
Danon \etal{} \cite{ref:danon} suggested that a normalized
variant ~\cite{ref:fredjain} of mutual information %of \eqnref{eq:mi}
be adapted for use in evaluating similar community partitions. 
Using \eqnarefs{eq:shannonentropy}{eq:mi}, 
the normalized mutual information $I_N (A,B)$ 
between partitions $A$ and $B$ is %defined as
\begin{equation} \label{eq:nmi}
  I_N(A,B) = \frac{ 2I(A,B) }{ H(A) + H(B) }
\end{equation}
which can take values in the range $0\le I_N (A,B)\le 1$.
Fred and Jain \cite{ref:fredjain} introduced, for computer vision problems,
a single resolution application of NMI that we use in our work.

The variation of information~\cite{ref:vi}
is a metric in the formal sense of the term and measures
the ``distance'' in information between two partitions $A$ and $B$.
Using \eqnarefs{eq:shannonentropy}{eq:mi}, 
$V(A,B)$ is calculated by 
\begin{equation} \label{eq:vi}
  V(A,B) = H(A) + H(B) - 2I(A,B).
\end{equation}
As an information distance, low values of $V(A,B)$ 
indicate better agreement between partitions $A$ and $B$.
VI has a range $0\le V(A,B)\le\log {N}$.
It is sufficient and even preferable to use the 
un-normalized version of VI.
We utilize both NMI and VI to demonstrate that our approach 
is not limited to a specific measure.

The mutual information $I$ and Shannon entropy $H$ 
also play a supplemental role in determining 
multiresolution structure.
For the Shannon entropy $H$, we average over all replicas using 
\begin{equation} \label{eq:averageH}
\langle H \rangle = \frac{1}{r}\sum_A H(A).
\end{equation}
For $I_N$, $V$, and $I$, we calculate the average of the 
measures over all pairs of replicas with 
\begin{equation} \label{eq:averageS}
  \langle S \rangle = \frac{2}{r(r-1)}\sum_{A>B} S(A,B)
\end{equation} 
where $S$ is any of the information measures
and $r$ is the number of replicas.
We use base $2$ logarithms in all information calculations.

Similarly, higher order cumulants of $S$ can be computed
with a (replica symmetrically weighted) probability 
distribution function that we set to be
\begin{equation}
  P(S) = \frac{2}{r(r-1)} \sum_{A > B} \delta\big[S-S(A,B)\big].
  \label{eq:repa}
\end{equation}
In \eqnref{eq:repa}, 
$\delta\big[S-S(A,B)\big]$ is the Dirac delta function.
For any function $f$ of $S$, the expectation value
of $f$ is %given by
\begin{equation}
  \langle f \rangle = \int dS ~P(S) f(S).
\end{equation}
Formally, in our probability distribution of \eqnref{eq:repa},
the information measure
$S$ plays a role analogous to the overlap parameter
in spin-glass problems.

\section{Potts Model Hamiltonian}
\label{sec:model}

We briefly review our Potts model approach to community 
detection~\cite{ref:rz}.
Generally speaking, community detection algorithms based 
on quality functions begin a community evaluation by measuring 
the number of connected nodes within (or outside) proposed 
communities.
In general, these edges can be weighted or unweighted.
The quality function must contrast this measure to some 
``expected'' value or directly evaluate missing connections 
in some manner.
If a linear addition of edge weights (connected and unconnected) 
is applied, 
the constructed model is equivalent to a Potts model spin system.
As it applies to community detection, 
such a model was first proposed by RB~\cite{ref:smcd}, 
which demonstrated a clear bridge between community 
detection methods and statistical physics.
RB implemented their model with a weighted comparison 
to a random partition (a ``null'' model) which included modularity 
as a special case.

In our Potts model, we directly sum the edge weights 
(connected and unconnected) in an energy calculation 
without a weighted null model.
Thus we avoid a comparison to the properties of another graph, 
random or otherwise, and despite a global energy sum, 
we obtain an effectively local measure of community structure.
As a local measure, the model is also free from the resolution 
limit discussed in the 
literature~\cite{ref:resolutionlimitdef,ref:fortunato,ref:kumpula}.

The general weighted Hamiltonian for our model is
\begin{equation}
	\Ham(\{ \sigma \}) = - \frac{1}{2} \sum_{i\neq j}
	\left( a_{ij} A_{ij} - \gamma b_{ij} J_{ij} \right)  
	       \delta (\sigma_i,\sigma_j)
	\label{eq:pottsmodel}
\end{equation}
which we refer to as an ``absolute Potts model'' (APM).
$A_{ij}= 1$ if nodes $i$ and $j$ are connected and are 
$0$ otherwise.  $J_{ij} \equiv (1-A_{ij})$.
The values $\{a_{ij}\}$ and $\{b_{ij}\}$ are general positive 
weights of the connected and unconnected edges, respectively,
which allow both symmetric and directed graphs.
$\{A_{ij}\}$, $\{J_{ij}\}$, $\{a_{ij}\}$, and $\{b_{ij}\}$ 
are all fixed by the definition of the system.
$\gamma$ is an externally defined weighting parameter 
for the unconnected edge weights.
In practice, we use a symmetric matrix with integer weights 
(faster integer computations)
on both connected and unconnected edges 
($\gamma$ is a rational number).
$\sigma_{i}$ is a Potts spin variable that can take 
an integer value $1 \le \sigma_{i} \le q$.
The value of $\sigma_{i}$ for a given node is the 
model equivalent of community membership.
That is, node $i$ is a member of community $k$ if $\sigma_{i}=k$.
The number of spin states $q$ can be specified as a constraint 
or can be determined by the lowest energy configuration 
over all values of $q$.
The Kroneker delta $\delta (\sigma_i,\sigma_j) = 1$ 
if $\sigma_i = \sigma_j$ and $\delta (\sigma_i,\sigma_j) = 0$ 
for $\sigma_{i} \neq \sigma_{j}$.
As in~\cite{ref:gudkov,ref:gudkovclique}, the interaction 
between spins is attractive if they are connected and 
repulsive if they are not connected. 
A further important feature of the Hamiltonian 
%of \eqnref{eq:pottsmodel} 
is that each spin interacts only with other spins 
in the same community.
The optimal ground state of \eqnref{eq:pottsmodel} is 
often difficult to locate in practice, 
so we identify the communities of a system by searching 
for low-energy states of this Hamiltonian.
%We will refer to this Hamiltonian as an 
%``absolute Potts model" (APM).

The edge density of a particular community $k$ is 
$p_k = 2l/[n(n-1)]$ where $l$ is the number of edges 
and $n$ is the number of nodes in the community.
We can relate the model weight parameter $\gamma$ 
to the \emph{minimum} internal edge density $p_{in}$ 
for every community. 
We obtain this relation from a simple calculation 
on the minimum number of interior edges that results 
in an energy of zero or less for a single community. 
An alternative method is to calculate the minimum number 
of edges that will merge two connected communities. 
Then we can apply an inductive argument to establish 
the same inequality.
For unweighted graphs, the relation is 
\begin{equation} \label{eq:minp}
  p_{in}\ge \frac{\gamma}{\gamma +1}.
\end{equation}
For a weighted graph, the relation is similar,
%$p_{in}\ge \gamma/[(\gamma +\overline{w})\overline{w}]$
$p_{in}\ge \gamma/(\gamma +\overline{w})$,
where $\overline{w}$ is the average weight of connected 
edges within each community and $p_{in}$ is then the edge 
weight density as compared to a maximally connected 
community with the same average weight $\overline{w}$.
These density relations are useful because the typical 
internal community edge density is equivalent 
to the \emph{resolution} of a system.
As a result, the resolution for the graph as a whole 
is also effectively set by $\gamma$.
This property is distinct from the resolution limit 
in the literature because the resolution set by this 
method is independent of a graph's own global 
parameters~\cite{ref:fortunato,ref:kumpula,ref:resolutionlimitdef}.

\section{Community Detection Algorithm}
\label{sec:cdalgorithm}

We apply the Potts model of \eqnref{eq:pottsmodel} 
with a simple community detection algorithm 
that is nevertheless extremely accurate, 
at least as accurate as the best available algorithms
(see Appendix \ref{app:cdaccuracy}) 
when used with our model~\cite{ref:rz}.
The algorithm sequentially ``picks up'' each node and places it 
in the community that best lowers the energy based 
on the current state of the system.
We repeat this process for all nodes and continue iterating
until no moves are found after one full cycle through all nodes.
This part of the dynamical approach is similar to parts 
of algorithms used in~\cite{ref:LPA,ref:blondel}.
We can also choose to test the communities 
for possible merges that can arise due to local minima 
traps~\cite{ref:clustermergenote}.
This test is more important for heavily weighted 
graphs with $\gamma\ll 1$.
We can optionally further allow zero-energy moves 
for difficult problems.
We attempt $t$ independent optimization trials 
(generally \bigO{1}) and select the lowest energy 
configuration as the solution.
Each trial permutes the order in which the nodes are 
initially traversed.
Appendix \ref{app:cdtransition} illustrates the effect 
of additional trials using a common benchmark problem 
with increasing levels of noise.
%This last step mitigates the effect of local energy 
%minima traps.

The algorithm has been modified to use the intuitive 
neighbor-node search such as 
in~\cite{ref:clausetlarge,ref:LPA,ref:blondel} and 
a symmetric initial state of one node per cluster also 
in~\cite{ref:clausetlarge,ref:LPA,ref:blondel} 
and applied in a more general dynamical context 
in~\cite{ref:gudkov}.
We further optimize the algorithm by allowing it 
to skip nodes that are already strongly defined 
within their respective communities.
Empirically, we find that the neighbor search 
drastically improves performance for sparse graphs 
to \bigO{N^{1+\beta}Z^{1+\beta}t\log Z} for some small 
$\beta$~\cite{ref:betadef,ref:logznote}, where $N$ is 
number of nodes and $Z$ is the average node degree. 
%of all nodes.
%and $t$ is the number of optimization trials attempted
The factor of $\log Z$ is due to a neighbor-node binary search 
for each connection matrix ($A_{ij}$ or $J_{ij}$) evaluation.
The factor of $NZ$ is due to the iteration over all neighbors,
and the factor of $\beta$ in the exponent is 
%and $\beta$ in the exponent 
due to the number of full $NZ$ iterations 
%which depends on the topology and the initial state of the system.
which depends on the topology of the system, the initial state 
of the system, and the resolution being solved
(\ie{}, the model weight $\gamma$).
This scaling enables us to achieve systems of at least 
\bigO{10^7} nodes and \bigO{10^9} edges for a single application 
of the algorithm.
Details of one of the large tests are discussed in Appendix \ref{app:cdlarge}.
We have solved systems up to \bigO{10^5} nodes and \bigO{10^7} 
edges for the multiresolution algorithm~\cite{ref:computerused} 
as discussed in \secref{sec:largehierarchy}.

\section{Multiresolution Algorithm}
\label{sec:algorithm}

One challenge in developing a multiresolution algorithm 
is that of selecting the best resolution(s) for the system.
A straight-forward method that avoids the choice 
of resolution is to iteratively solve the system 
(with a necessary change in $\gamma$ for our model) 
and collapse the communities into ``supernodes'' until 
the system is organized into a forced hierarchical structure.
This approach is viable; 
but even when the system is hierarchical in nature,
there is the question of whether the best resolutions
were resolved at each stage.
Our algorithm enables a quantitative analysis that 
determines the best resolutions and applies to general 
types of multiresolution structure.

\subsection{Motivation}

Ideally, we desire an algorithm that allows the system 
to communicate what the best resolutions are;
but without \emph{a priori} information, the correct weights 
for these resolutions are not obvious in general.
In order to identify the proper resolutions, 
we examine information-based correlations among independent 
replicas (independent solutions) via NMI or VI over a range 
of resolutions.
Rather than using the replicas to simply identify a unique 
optimized solution for each resolution, 
we examine correlations among the entire set. 
We then select the strongest correlations as the best 
resolutions.
%These correlations allow us to discern information about 
%the system beyond simply an optimized solution.

From a global perspective, the average NMI 
(between all pairs of replicas) indicates 
how strongly a given structure dominates the energy landscape 
by measuring how well the replicas agree with each other.
High values of the NMI (often manifested as peaks)
correspond to more dominant, 
and thus more significant, structures.
From a local perspective, at resolutions where the system 
has well-defined structure, a set of independent replicas 
should be highly correlated because the individual nodes 
have strongly preferred community memberships.
Conversely, for resolutions ``in-between'' two strongly defined 
configurations, 
one might expect that independent replicas will be less 
correlated due to ``mixing'' between competing divisions of the graph.
Random effects will usually reduce the correlations 
between independent solutions.

A similar argument applies to VI where, 
as an information distance, low values of VI
correspond to better agreement among replicas.
With these information-based correlations, 
we obtain a set of multiresolution partitions of the graph, 
but we also obtain an estimate of the relative strength
of the structures at each resolution.
Note that this argument does not distinguish between 
unrelated multiresolution structures or those that 
are strictly hierarchical in nature although nothing prevents 
the imposition of additional hierarchical constraints if desired.

Implicit in this argument is the idea that local minima
in the energy landscape represent meaningful, 
even if perhaps incomplete, information about the graph.
The same assertion was made in~\cite{ref:salespardo,ref:reichardt}
for modularity and the RB Potts model.
Moderate levels of ``confusion'' caused by random 
or competing effects within a graph do not destroy 
information contained in the global energy landscape,
and the replica correlations of our algorithm are 
a measure of the ``complexity'' of that landscape.
As the noise in the system is increased we expect that 
the transition to incoherence 
(where replicas are weakly correlated) to occur rapidly
(see end of \secref{sec:discussion}
and a brief example of an accuracy transition 
in Appendix \ref{app:cdtransition}).
%Such behavior for random graphs is reviewed in~\cite{ref:albert}.
If an algorithm can verifiably solve for the global 
minima of a system in most cases, 
the problem of community detection is solved in principle.
Since this is difficult to do in practice, the replica 
correlations in our algorithm take advantage of the fact 
that we cannot always locate the optimal ground state(s).

In principle, one can also include in \eqnref{eq:pottsmodel}
interactions between each of the $r$ replicas to produce 
a ``free energy'' type functional of the form 
\begin{equation}
	F = \sum_i \Ham_i(\{ \sigma \}) - T \sum_{i\neq j} S(i,j).
	\label{eq:replicainteraction}
\end{equation}
where $S(i,j)$ is an information-based measure 
(e.g., $I_N$, $V$, etc.) between all replica pairs
and $T$ is a scale for this information measure. 
$S(i,j)$ is maximized when the community partitions 
are identical in all replicas. 
This information theory measure formally plays a role 
analogous to entropy in a free energy functional.
%As the information measure scale factor, 
$T$ then plays the role of a ``temperature.''
Sans the first term, the minima of $F$ 
in \eqnref{eq:replicainteraction} produce highly 
correlated random configurations 
(a ``random high temperature configuration'' of the system 
which appears without change in all replicas).
Our algorithm in this work will amount to initially 
minimizing the first term in $F$, 
\ie{}, $\sum_i \Ham_i(\{ \sigma \})$,
for a set of fixed $\{\gamma_i\}$.
Out of this set of replica configurations,
we then ask for which $\gamma_i$ do we find 
a maximum of the correlations, $\sum_{i\neq j} S(i,j)$, 
when this information theory measure is plotted 
as a function of $\gamma$.  
A more sophisticated version of our algorithm minimizes 
$F$ directly with both terms included in each step.
%(see end of \secref{sec:discussion}).  
The information theory measures that we employ 
%to determine the correlation between different replicas 
may also be written for other (non-graph theoretic) 
optimization problems with general Hamiltonians, 
or cost functions, $\cal{H}$
(see Appendix \ref{app:generalentropy}).

\subsection{Algorithm}

We start the algorithm with a weighted or unweighted graph.
In \eqnref{eq:minp}, $p_{in}$ is the \emph{minimum} internal 
edge density for each community, and it is equivalent 
to the \emph{resolution} of the system when we 
minimize \eqnref{eq:pottsmodel}.
The algorithm uses \eqnref{eq:pottsmodel}
to solve a range of resolutions $\{p_i\} = [p_{0},p_{f}]$
(decrementing $p_i$)
corresponding to a particular set of model weights 
$\{\gamma_i\} = [\gamma_{0},\gamma_{f}]$ 
as determined by \eqnref{eq:minp}.
It is almost always sufficient 
to have $\gamma_0\lesssim 19$ since it corresponds 
to a \emph{minimum} community edge density of $p_{0}\ge 0.95$.
The final weight $\gamma_f$ is found
when the system is completely reduced.
A completely reduced system is one that is fully collapsed 
into one community or one where disjoint sub-graphs 
will not allow the system to collapse any further.

Each iteration, we decrement the density $p_i$ 
by a small value $\Delta p = 0.05$ 
(or $0.025$ for smaller graphs)
and calculate the corresponding $\gamma_i$.
After a threshold value (say $p_t = 0.1$),
we scale $p_i$ by a factor of $1/2$
(or $3/4$ for smaller graphs)
in order to take sizable steps towards a fully 
reduced system (necessary for large systems).
One could readily implement an adaptable step 
or ``fill-in'' process since the order of trials 
is irrelevant for the result.

The algorithm takes three input parameters:
the number of independent replicas $r$ that will be 
solved at each tested resolution,
the number of trials per replica $t$,
and the starting density which we set to be 
$p_0\simeq 0.95$ corresponding to $\gamma_0 = 19$.
The number of replicas is typically $8\le r\le 12$
and is selected based upon how much averaging 
(over all replica pairs) is needed or desired.
The number of trials $t$ per replica is generally 
$2\le t\le 20$.
For each replica, we select the lowest energy solution 
among the $t$ trials as was discussed 
in \secref{sec:cdalgorithm}.
The value of $t$ is chosen based on how much optimization
is necessary to identify a strong low-energy 
configuration~\cite{ref:clustermergenote}.

The $r$ replicas (and $t$ optimization trials) are 
generated by reordering the ``symmetric'' initialized 
state of one node per community.
That is, even though the initialized state is symmetric,
the order that we traverse the list also affects
the answer that we obtain.
This occurs because the node-level dynamics of the underlying 
community detection algorithm in \secref{sec:cdalgorithm} 
moves a node immediately upon identifying the best community 
membership given the current state of the system.
Utilizing the $r$ replicas, we then use the 
%information correlation among the $r$ replicas 
information-based measures of \secref{sec:informationmeasures} 
to determine the multiresolution structure. 
%In the following steps, 
%$N$ is the number of nodes in the graph,
%$n$ is the average number of nodes in all communities,
%and $q$ is the number of communities.
Our algorithm is given by the following steps:
\begin{enumerate}
     \item \emph{Initialize the system}.  Initialize adjacency
     matrices ($A_{ij}$ and $J_{ij}$) and weights 
     ($a_{ij}$ and $b_{ij}$) based on the system definition.
     Use \eqnref{eq:minp} and $p_0$ to calculate the initial
     model weight $\gamma_0$. \label{step:init}

     \item \emph{Solve all replicas at this resolution} $p_i$.
     Initialize the current replica to a 
     symmetric state of one node per community.
     Use \eqnref{eq:pottsmodel} to solve each replica
     with model weight $\gamma_i$ 
     at a cost of \bigO{N^{1+\beta}Z^{1+\beta}t\log Z} 
     per replica~\cite{ref:clustermergenote,ref:betadef}.
     Repeat the process independently for all $r$ replicas.
     Each trial and replica randomly permutes the order 
     in which nodes are initially traversed 
     in the respective solutions.
     \label{step:solutions}

     \item \emph{Calculate the replica} $I_N$, $V$, $I$, \emph{and} $H$ 
     \emph{information measures}.
     Use \eqnref{eq:shannonentropy} to calculate $H$ for all 
     replicas and \eqnrefs{eq:mi}{eq:vi} to calculate $I$, $I_N$, 
     and $V$ between all pairs of replicas
     for this resolution $p_i$~\cite{ref:nmicalcnote}.
     Calculate the average (see \eqnarefs{eq:averageH}{eq:averageS}) 
     and the standard deviation for each 
     measure. \label{step:correlations}

     \item \emph{Decrement to the next resolution} $p_{i+1}$. 
     If $p_i> 0.1$, decrement $p_{i+1} = p_i-0.05$ 
     or $0.025$ for smaller graphs.
     If $p_i\le 0.1$, $p_{i+1} = p_i/2$
     or $3p_i/4$ for smaller graphs.
     Calculate the model weight 
     $\gamma_{i+1}$ by \eqnref{eq:minp}.
     Return to Step~\ref{step:solutions} until the system 
     is not further reducible (fully collapsed or disjoint 
     sub-graphs will not collapse). \label{step:decrement}
     
	   \item \emph{Evaluate results}.  
	   For the range of model weights $\{\gamma_i\}$, 
     plot each average $I_{N,i}$, $V_i$, $I_i$, and $H_i$ 
     versus $\gamma_i$.  Determine the strongest correlations 
	   ($I_N$ high or $V$ low) in these plots 
	   (see Figs.\ \ref{fig:hierarchyplot} -- \ref{fig:largehierarchyplot},
	   \ref{fig:dolphinsplot}, \ref{fig:highlandplot}, 
	   and \ref{fig:lfksampleplot}).  
	   These strongly correlated regions correspond 
	   to the best multiresolution structure(s) in the graph.
	   If the correlation is less than ``perfect'' 
	   ($I_N < 1$ and $V>0$),
	   we choose the lowest energy replica to be the partition solution.
	   One could also choose to construct a ``consensus" partition
	   between all of the replicas~\cite{ref:LPA,ref:fredjain} 
	   at each notable resolution.
\end{enumerate}
We estimate that the number of resolutions $\{p_i\}$ required 
to adequately specify an arbitrary system scales as \bigO{\log N}.
The dominant scaling of the algorithm is almost always 
Step~\ref{step:solutions}, so we estimate that the 
overall scaling is \bigO{N^{1+\beta}Z^{1+\beta}rt\log N\log Z} 
for some small $\beta$~\cite{ref:betadef,ref:logznote}.

% *** hierarchy figures ***************************************
% *** Placed here to get 19 page formatting *******************
\myfig{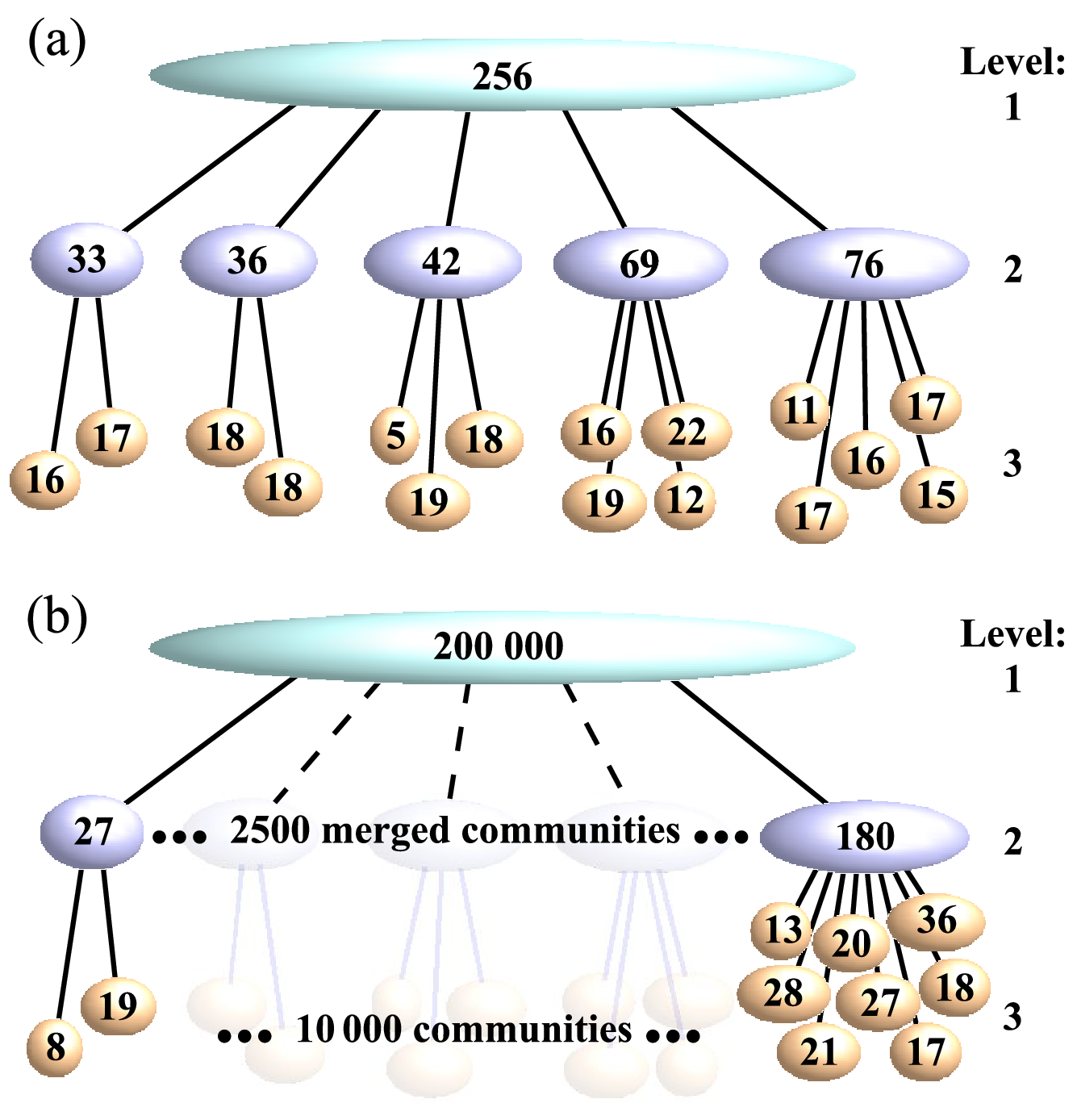}{
(Color online) 
Heterogeneous hierarchical systems corresponding 
to the plots in \figref{fig:hierarchyplot} for panel (a)
and the plots in \figref{fig:largehierarchyplot} for panel (b).
In panel (a), 
the $256$ node system is divided into a three-level hierarchy where 
the unweighted edge connection probabilities at each level are
the following:
level $3$ has $p_3=0.9$ between nodes in the \emph{same} community
with community sizes from $5$ to $22$ nodes (average $16$).
Level $2$ has $p_2=0.3$ between nodes in \emph{different} constituent 
sub-communities with merged community sizes from $33$ to $76$ nodes.
Level $1$ is the completely merged system of $256$ nodes 
with $p_1=0.1$ between nodes in \emph{different} sub-communities.
The average edge density is $p=\overline{p}_1=0.182$.
In panel (b), we increase the system size to $200~\!000$ nodes.
Level $3$ has $10~\!000$ communities
with sizes from $6$ to $37$ nodes (average $20$).
Level $2$ has $2500$ communities 
with sizes from $27$ to $180$ nodes which are formed by 
merging two to eight communities from level $3$.
The density $p_1$ is changed from panel (a) to $p_1=0.000~\!31$,
and the average edge density is $p=\overline{p}_1\simeq 0.0005$.
This larger system has over ten million edges 
with approximately $62\%$ of the edges being random 
noise between level $2$ communities.
}{fig:heterohierarchy}{0.9\linewidth}{t}

\myfig{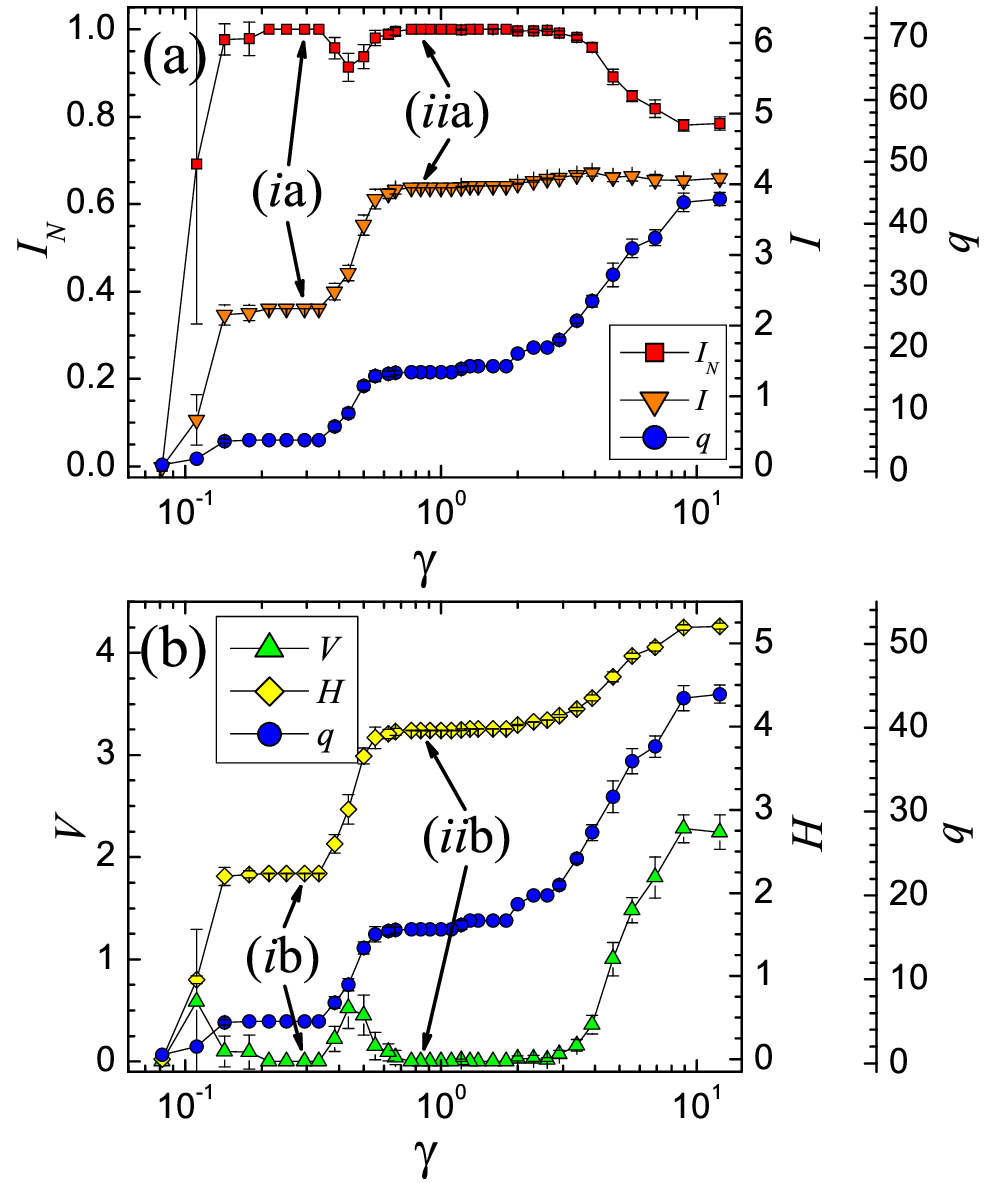}{
(Color online) 
Plot of information measures $I_N$, $V$, $H$, and $I$ 
in panels (a) and (b) vs. the Potts model weight 
$\gamma$ in \eqnref{eq:pottsmodel} for the three-level 
heterogeneous hierarchy depicted 
in \subfigref{fig:heterohierarchy}{a}.
In panel (a), the squares represent the average replica 
normalized mutual information $I_N$ (left axis), and the inverted triangles 
represent the average mutual information $I$ (right axis).
In panel (b), the triangles represent the average 
variation of information $V$ (left axis), 
and the diamonds represent the average 
Shannon entropy $H$ (right axis).
For comparison, the circles in both panels (a) and (b) 
represent the average number of clusters $q$ for the same 
set of replicas (right-offset axes). 
In panel (a) the peak $I_N$ values ($i$a) and ($ii$a) 
both accurately correspond to levels $2$ and $3$ respectively 
of the hierarchy depicted in \subfigref{fig:heterohierarchy}{a}.
Similarly in panel (b) the minimum $V$ values ($i$b) and ($ii$b) 
also accurately correspond to levels $2$ and $3$, respectively,
of the hierarchy.
In panels (a) and (b), both the mutual information $I$ 
and Shannon entropy $H$ display a ``plateau" behavior 
corresponding to the correct solutions.
Plateaus in the average number of clusters 
$q$~\cite{ref:qaveragenote} also
indicate important structures as in~\cite{ref:multires}.
}{fig:hierarchyplot}{3.0in}{t}
% *** end hierarchy figures ************************************

Structures identified by this algorithm are not necessarily 
hierarchical; however, one can augment the algorithm 
by imposing an additional hierarchical constraint
on some fraction of the replicas.
Comparisons would then be made strictly between all pairs 
with and without this additional constraint.
We applied this variation in both divisive and agglomerative 
approaches, but in our testing it only resulted in a modest 
improvement to the algorithm's ability to identify the best 
resolutions.
Therefore, we use the above algorithm in order to
take advantage of its generality and relative simplicity.

\section{Examples}
\label{sec:examples}

We show the results of the multiresolution algorithm 
of \secref{sec:algorithm} applied to several test 
cases~\cite{ref:datawebsite}.
In \secsref{sec:smallhierarchy}{sec:largehierarchy}, 
we illustrate a small $256$ node and a larger 
$200~\!000$ node hierarchy respectively with 
both systems depicted in \figref{fig:heterohierarchy}.
In \secref{sec:randomgraph}, we examine the structure
of an Erd\H os-R\'enyi random graph 
for comparison to graphs with known internal structure.
We then analyze two real social networks
in \secsref{sec:dolphins}{sec:highland}
where the respective systems are depicted in \figsref{fig:dolphinsgraph}{fig:highlandgraph}.
In \secref{sec:accuracy}, we also demonstrate the 
algorithm's exceptional accuracy for large systems.

% *** hierarchy discussion ************************************

\subsection{256 node hierarchy} \label{sec:smallhierarchy}

The system in \subfigref{fig:heterohierarchy}{a} 
depicts a set of $256$ nodes for a 
constructed three-level heterogeneously-sized hierarchy. 
The results are seen in \figref{fig:hierarchyplot}.
The unweighted edge connection probabilities 
are $p_k$ for $k=1,2,3$. Level $3$ %at the bottom 
has a density $p_3=0.9$ between nodes 
in the \emph{same} community with community sizes from $5$ to $22$ 
(average $16$) nodes.  Level $2$ %in the middle 
has a density $p_2=0.3$ between nodes 
in \emph{different} constituent sub-communities and is divided 
into five groups with merged sizes from $33$ to $76$ nodes.
Level $1$ is the completely merged system that has a density 
$p_1=0.1$ between nodes in \emph{different} sub-communities.
These edges provide some system noise.
The average densities of communities at levels $1$ and $2$ are
$p = \overline{p}_1=0.182$ and $\overline{p}_2 = 0.470$.
We use eight replicas and four trials per replica
at a total run time of $6.1$ s~\cite{ref:computerused}.

In \subfigref{fig:hierarchyplot}{a}, the squares represent 
NMI averages over all replica pairs (left axis). 
The inverted triangles represent the mutual information $I$ 
averages for the same replica pairs (right axis).
In \subfigref{fig:hierarchyplot}{b}, 
the triangles represent VI averages over all 
replica pairs (left axis),
and the diamonds represent the Shannon entropy $H$
averages for the replicas (right axis).
In both panels, the circles represent the average number
of clusters across the replicas (right offset axes).
All parameters are plotted versus the model weight 
$\gamma$ where we use a logarithmic scale to facilitate 
comparing the behavior of a large range of system 
sizes from $N=16$ nodes 
in \figsref{fig:highlandgraph}{fig:highlandplot}
to as large as $N=200~\!000$ nodes 
in Figs.\ \ref{fig:heterohierarchy}(b) 
and~\ref{fig:largehierarchyplot}~\cite{ref:logscalenote}.

The extrema ($i$a,b) and ($ii$a,b) are the 
correctly determined levels $2$ and $3$ respectively
of the test hierarchy %that are 
depicted in \subfigref{fig:heterohierarchy}{a}.
Peaks ($i$a) and ($ii$a) have $I_N = 1$ 
and minima ($i$b) and ($ii$b) have $V = 0$
which indicate perfect correlations among the replicas
for both levels of the hierarchy.
The ``plateaus'' in $H$ and $I$ are a second indication
of the significant system structure whose importance will 
become more apparent in later examples.
The plateau in the average $q$~\cite{ref:qaveragenote} 
is also an important indicator of system
structure as used in~\cite{ref:multires}.
However, Figs.\ \ref{fig:randomhierarchyplot}, 
\ref{fig:dolphinsplot}, and \ref{fig:highlandplot}
discussed later demonstrate that some caution should be 
exercised when using the plateau criterion (in $H$, $I$, or $q$)
for determining multiresolution structure.

At level $3$ in \subfigref{fig:heterohierarchy}{a}, 
the average number of externally connected 
edges for each node is $Z_{out}\simeq 32.0$
with a random noise component of $Z_{out}^{noise}\simeq 19.8$.
Both of these numbers are larger than the average number 
of internal edges, $Z_{in}\simeq 14.3$.
Despite this imbalance, the algorithm easily identifies 
level $3$ of the hierarchy because the external edges 
(particularly those due to the random noise) are not 
concentrated strongly enough into any one external cluster.
This behavior is important for smaller 
communities on level $3$ 
where $Z_{out}$ is substantially larger than $Z_{in}$,
and it illustrates that the model is robust to noise 
in the system.

The VI peaks at $\gamma_1= 0.111$ and $\gamma_2= 0.435$
in \subfigref{fig:hierarchyplot}{b} correspond to the 
average inter-community edge densities, $p_1=0.1$ for 
sub-communities at level $2$ and $p_2=0.3$ 
for sub-communities at level $3$.
Equation (\ref{eq:minp}) relates the \emph{minimum} internal 
edge density $p_{in}\ge\gamma/(\gamma+1)$
for each community in a solved partition.
We can arrive at this inequality, using inductive 
reasoning, by considering the minimum inter-community 
edge density required for two arbitrary communities 
$A$ and $B$ to merge.
We apply the relation %of \eqnref{eq:minp} 
as an equality (\ie{}, energy difference between the merged 
and unmerged states is approximately zero)
for the peak VI values at $\gamma_1$ and $\gamma_2$.
The respective densities are $p_1^{AB}= 0.100$ 
and $p_2^{AB}= 0.303$.
These values correspond closely to the constructed
inter-community densities $p_1$ and $p_2$ above.
The local VI maxima show that ``complexity" 
of the energy landscape increases at resolutions 
where $\gamma/(\gamma+1)$ %in \eqnref{eq:pottsmodel} 
is equal to the mean inter-community edge density.
The more intuitive interpretation is that the  
``complexity" of the energy landscape 
increases substantially when the energy difference 
between different states is approximately zero.

% *** end hierarchy discussion ************************************

% *** Random graph discussion *************************************

\subsection{Erd\H os-R\'enyi random graph}  \label{sec:randomgraph}

% *** Random graph figure *****************************************

\myfig{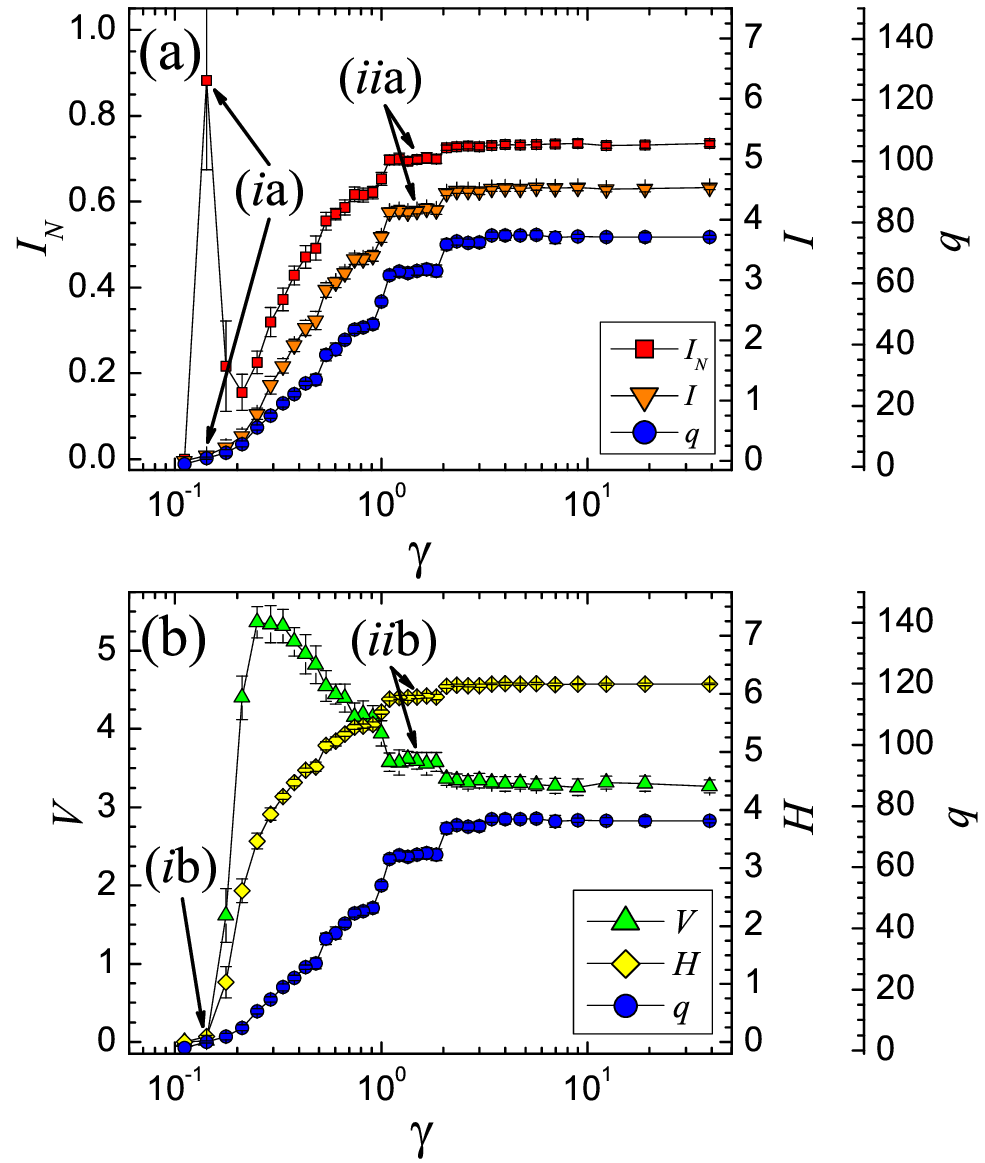}{
(Color online) 
Plot of information measures $I_N$, $V$, $H$, and $I$ in panels
(a) and (b) vs the Potts model weight $\gamma$ 
for a purely (Erd\H os-R\'enyi) random graph that 
has the same average density $p=0.182$ as
the hierarchy in \subfigref{fig:heterohierarchy}{a}
and the corresponding results in \figref{fig:hierarchyplot}.
The right-offset axes plot the number of clusters $q$.
See \figref{fig:hierarchyplot} for a complete description
of the legends and axes.
In panel (a), the peak ($i$a) corresponds to a 
trivial partition of the system into groups with sizes 
of approximately $\{1,2,253\}$ among the different replicas.
The trivial structure change in the NMI spike is indicated
by its the low value of mutual information $I$ at ($i$a) 
and by its low VI $V$ and Shannon entropy $H$ at ($i$b).
The plateaus at ($ii$a,b) do not correspond to a consistent 
multiresolution structure as evidenced by the poor 
NMI and VI correlations.
Rather, they indicate multiple similarly sized configurations
that have similar community edge densities.
}{fig:randomhierarchyplot}{3.1in}{t}

In \figref{fig:randomhierarchyplot},
for comparison purposes we show the results for a purely 
(Erd\H os-R\'enyi) random graph at the same average edge 
density $p=0.182$ as the hierarchy 
in Figs.\ \ref{fig:heterohierarchy}(a)
and~\ref{fig:hierarchyplot}.
We use eight replicas and four trials per replica
at a total run time of about $6.9$ sec~\cite{ref:computerused}.
The only peak ($i$a) in the random graph corresponds 
to a trivial division %of the graph 
into groups with sizes of approximately $\{1,2,253\}$ among
the various replica solutions.
This peak indicates transitional behavior to lower density, 
essentially trivial, structures.
Peaks such as ($i$) can be distinguished from more
meaningful ones by the cluster size distribution or 
the corresponding information measures.
The value of $I$ at ($i$a) or $V$ and $H$ at ($i$b)
all have very low information values.
Otherwise, the random graph displays no significant 
multiresolution structure.

All of the information measures display a plateau
behavior at ($ii$a,b).
The plateaus in NMI or VI do not indicate a clear 
multiresolution structure because the correlations are 
relatively poor ($I_N\simeq 0.70$ and $V\simeq 3.6$)
for both measures.
If we examine the detailed solutions across the plateaus
(separate from our multiresolution algorithm),
the average NMI and VI are $I_N=0.644$ and $V=4.04$
both of which indicate poor agreement.
There is no consistent structure identified by the 
community detection algorithm in this region.
Instead, the weak plateaus in NMI and VI indicate 
that the system is constrained within a set of 
similarly sized partitions that have similarly high 
community edge densities.
This example also illustrates that if we use \emph{only} 
the plateaus (in $H$, $I$, or $q$),
there is a potential to incorrectly 
identify significant structure(s) in the system.
This possibility can be remedied by information 
checks on nearby solutions in the plateau, but the poor 
NMI and VI correlations already appear to indicate the 
lack of consistent structure in the region.

% *** Large hierarchy figure *****************************************

\myfig{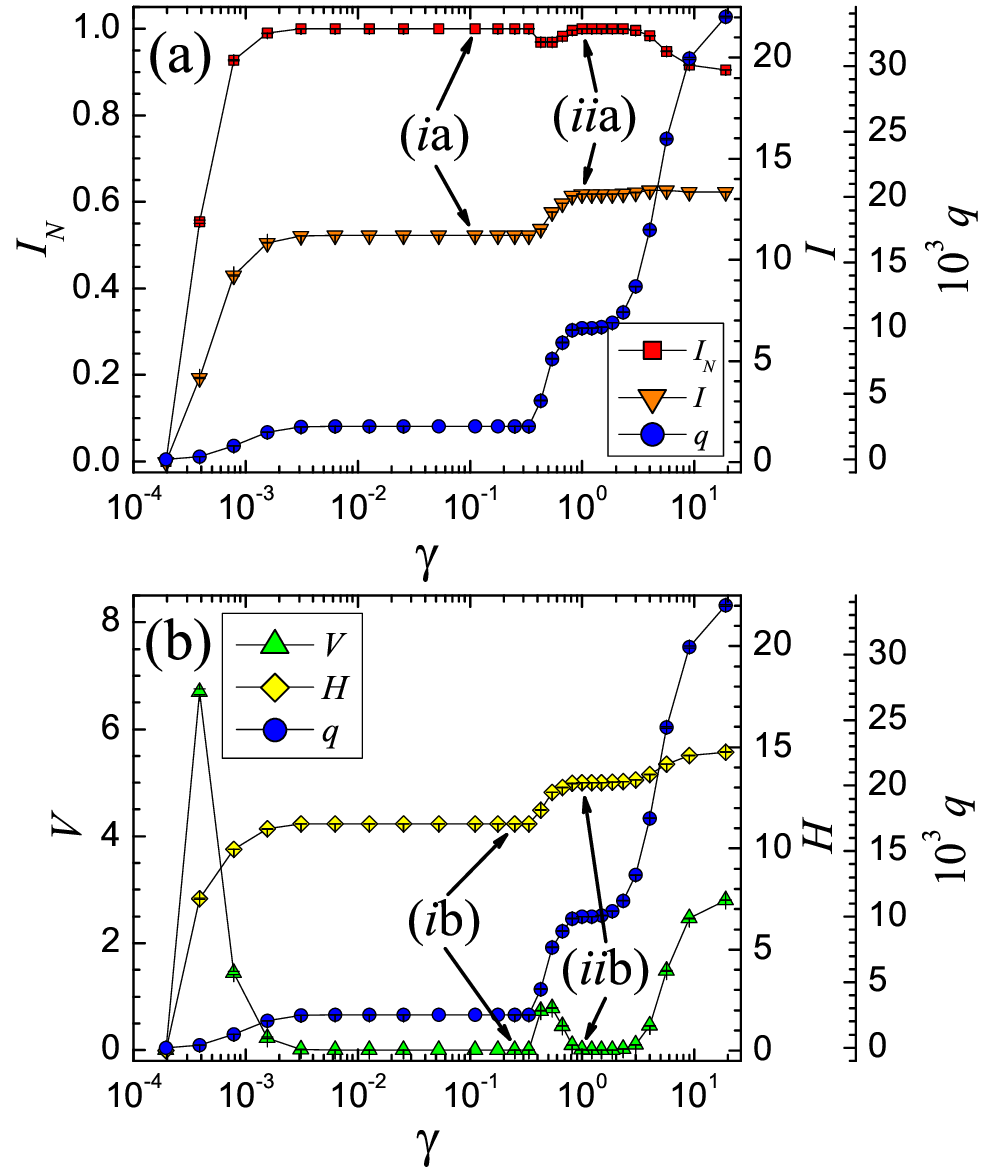}{
(Color online) 
Plot of information measures $I_N$, $V$, $H$, and $I$ 
in panels (a) and (b) vs. the Potts model weight 
$\gamma$ for the large three-level heterogeneous hierarchy 
depicted in \subfigref{fig:heterohierarchy}{b}.
The right-offset axes plot the number of clusters $q$.
See \figref{fig:hierarchyplot} for a complete description
of the legends and axes.
With the exception of $15$ weakly connected 
nodes (out of $200~\!000$) and $5$ merged clusters 
(out of $10~\!000$) at ($ii$a,b), 
the extremal values of $I_N$ and $V$ at ($i$a,b) and ($ii$a,b)
both accurately correspond to levels $2$ and $3$ respectively 
of the hierarchy depicted in \subfigref{fig:heterohierarchy}{b}. 
}{fig:largehierarchyplot}{3.1in}{t}

% *** Large hierarchy discussion *****************************************

\subsection{Large hierarchy}  \label{sec:largehierarchy}

A much larger hierarchy 
%than \subfigref{fig:heterohierarchy}{a}
is depicted in \subfigref{fig:heterohierarchy}{b}.
The system has $200~\!000$ nodes and $10~\! 011~\! 428$ edges.
Approximately $62\%$ of these edges are due to random noise 
between level $2$ communities.  For this system, 
$p_1=0.000~\!31$, but $p_2=0.3$ and $p_3=0.9$ are unchanged
from \subfigref{fig:heterohierarchy}{a}.
There are $10~\!000$ sub-communities at level $3$ with sizes 
ranging from $6$ to $37$.
Level $3$ communities are combined in groups of two to eight 
to form the $2500$ communities of level $2$ with sizes 
ranging from $27$ to $180$.
We use eight replicas and two trials per replica with a run 
time of about $4.6$ hours~\cite{ref:computerused}.
In \figref{fig:largehierarchyplot}, extrema ($i$a,b) exactly 
identify level $2$ of the hierarchy with perfect NMI and VI 
correlations, and extrema ($ii$a,b) accurately identify 
($I_N = 0.999~\!995$ and $V= 1.42\times 10^{-4}$)
all but $5$ merged clusters out of $10~\!000$ and $15$ nodes 
out of $200~\!000$ nodes for level $3$.
Due to random fluctuations, 
all of these nodes have a random connectedness of $50\%$ 
or less for their intended communities. 
This result is therefore consistent with the 
model and algorithm.

% *** Dolphin discussion *****************************************

\subsection{Dolphin social network}  \label{sec:dolphins}

% *** Dolphin figures *****************************************

\myfig{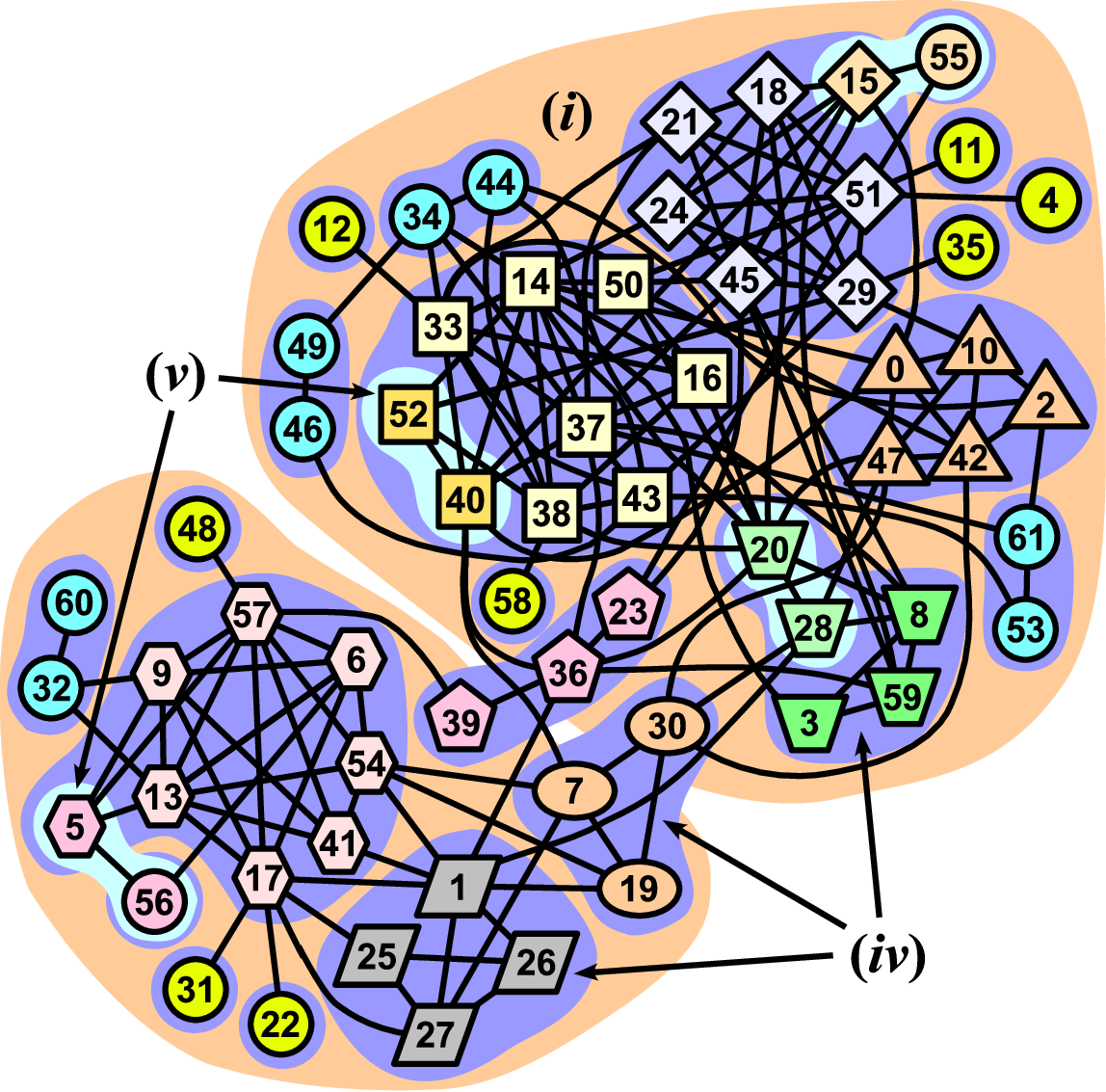}{(Color online) 
Pictorial representation of a social network of $62$ 
bottlenose dolphins in Doubtful Sound, New Zealand%
%Original data was collected by Lusseau
~\cite{ref:lusseausplit,ref:lusseaudata,ref:lusseauecol}.
These groupings correspond to structures 
($i$), ($iv$), and ($v$) in \figref{fig:dolphinsplot} 
in order of smaller group sizes.
The two-cluster partition ($i$) corresponds to a known 
split of the dolphin community~\cite{ref:lusseausplit}.
In partition ($iv$), sub-groups 
are assigned distinct node shapes except for 
circles which indicate various one and two member groups.
Structure ($v$) is identified from configuration ($iv$) 
when the four highlighted dyads of dolphins 
($\{5,56\}$, $\{15,55\}$, $\{20,28\}$, and $\{40,52\}$) 
form distinct sub-groups.
Note that sub-groups $\{7,19,30\}$ and $\{23,36,39\}$ 
in ($iv$) have nodes that are separated in their 
respective super-groups. 
These groups are examples of how our algorithm 
does not restrict node assignments between different 
resolutions, and they illustrate how the algorithm can 
apply to general types of multiresolution structure. 
}{fig:dolphinsgraph}{3.0in}{t}

\myfig{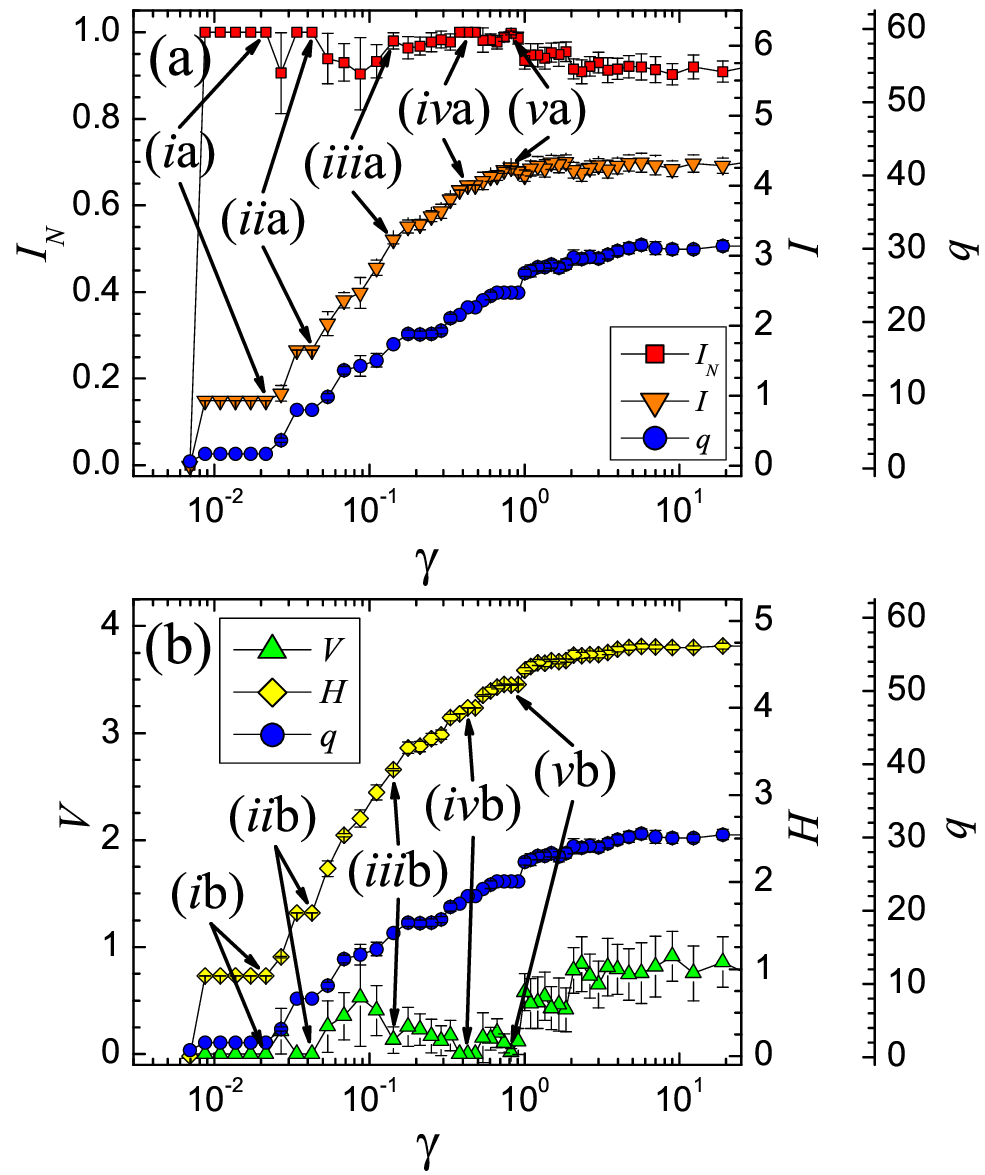}{(Color online) 
Plot of information measures $I_N$, $V$, $H$, and $I$ 
in panels (a) and (b) vs. the Potts model weight $\gamma$ 
for a social network of $62$ 
bottlenose dolphins in Doubtful Sound, New 
Zealand~\cite{ref:lusseaudata,ref:lusseausplit,ref:lusseauecol}.
A summary of results is depicted in \figref{fig:dolphinsgraph}
for configurations ($i$), ($iv$), and ($v$).
The right-offset axes plot the number of clusters $q$.
See \figref{fig:hierarchyplot} for a complete description
of the legends and axes.
One notable grouping is configuration ($i$) which corresponds 
to a known split of the dolphin community~\cite{ref:lusseausplit}.
The structures represented by ($ii$) -- ($v$) 
are other potential strongly defined partitions 
and are explained in the text.
}{fig:dolphinsplot}{3.0in}{t}

We tested a social network of $62$ bottlenose 
dolphins in Doubtful Sound, 
New Zealand~\cite{ref:lusseausplit,ref:lusseaudata,ref:lusseauecol}.
Three of the strongest partitions (($i$), ($iv$), and ($v$))
are depicted in \figref{fig:dolphinsgraph}
using the results in \figref{fig:dolphinsplot}.
We use ten replicas with ten trials per replica
at a total run time of about $0.78$ sec~\cite{ref:computerused}.
We use a density scaling of $0.8$ rather than $0.75$ 
for $p_{i}<0.1$ for Step~\ref{step:decrement} 
of the algorithm in order to more easily observe 
the transition between structures ($i$) and ($ii$) 
in \figref{fig:dolphinsplot}.
Configuration ($i$) identifies a grouping of $21$ and $41$
dolphins with perfect NMI and VI correlations ($I_N=1$ and $V=0$). 
This configuration agrees with an observed split of the 
dolphin network when a dolphin left the 
school~\cite{ref:lusseausplit}, but our algorithm also 
suggests that this configuration is not the only strongly 
defined partition for the system.

Our algorithm further identifies partitions ($ii$) -- ($v$) 
as important candidate partitions based on the strong 
NMI and VI information correlations.
Partition ($ii$) separates weakly connected dolphins 
($\{4\}$, $\{11\}$, $\{12\}$, $\{35\}$, $\{58\}$, 
and $\{46,59\}$) 
in the larger super-group of \figref{fig:dolphinsgraph} 
into distinct sub-groups.
Configuration ($iii$) is slightly less well-defined 
with information correlations of
$I_N\simeq 0.980$ and $V\simeq 0.132$.
It separates weakly connected dolphins
($\{22\}$, $\{31\}$, $\{39\}$, $\{48\}$, and $\{32,60\}$) 
of the smaller super-group of partition ($i$) 
and also begins a coarse division of the larger super-group.
Configuration ($iv$) is perfectly correlated and 
is the first major reconfiguration of both super-groups
of structure ($i$).
The data in the three largest groups of ($iv$) are 
largely divided along gender lines according to details 
presented in~\cite{ref:lusseaudata}.
Configuration ($v$) is a slight variation of ($iv$)
with $I_N\simeq 0.998$ and $V\simeq 0.0178$
which separates four dyads of dolphins
($\{15,55\}$, $\{46,49\}$, $\{32,60\}$, and $\{20,28\}$)
into distinct groups.
Among different tests, there is some variation in the 
predicted groupings where a few nodes can be reassigned 
between groups or separated into distinct communities.
Sub-groups $\{7,19,30\}$ and $\{23,36,39\}$ 
of configuration ($iv$) %in \figref{fig:dolphinsgraph} 
have nodes that are split between the two super-groups 
of ($i$). 
These groups show that our algorithm does not restrict 
node assignments between different resolutions.
This behavior allows our algorithm to solve general 
types of multiresolution structures. 

All measures show a strong plateau for configuration ($i$a,b).
The mutual information $I$ shows weak plateaus 
at ($ii$a) and ($iv$a) but no plateau at ($iii$a) and ($v$a). 
Similarly, the Shannon entropy $H$ shows weak plateaus 
at ($ii$b) and ($v$b) but no plateau for ($iii$b) and ($iv$b).
The average number of clusters $q$ as used 
in~\cite{ref:multires} also indicates the presence 
of structures ($ii$) and ($v$), %\figref{fig:dolphinsplot},
but it misses partition ($iv$).
Additionally, a weak plateau in $q$ near configuration ($iii$)
predicts a slightly different resolution than the extremal
NMI and VI correlations. 
The weak plateau behavior of $H$, $I$, or $q$ at 
different configurations of ($ii$a,b) -- ($v$a,b) 
do not contradict the existence of valid structures.  
Rather, missing plateaus in the supplemental measures 
$H$, $I$, or $q$ 
%which correspond to strong NMI and VI correlations 
can indicate a noisy graph in general 
or a strongly defined but transient resolution.

% *** Highland discussion *****************************************

\subsection{Highland Polopa tribe relations}  \label{sec:highland}

% *** Highland figures *****************************************

\myfig{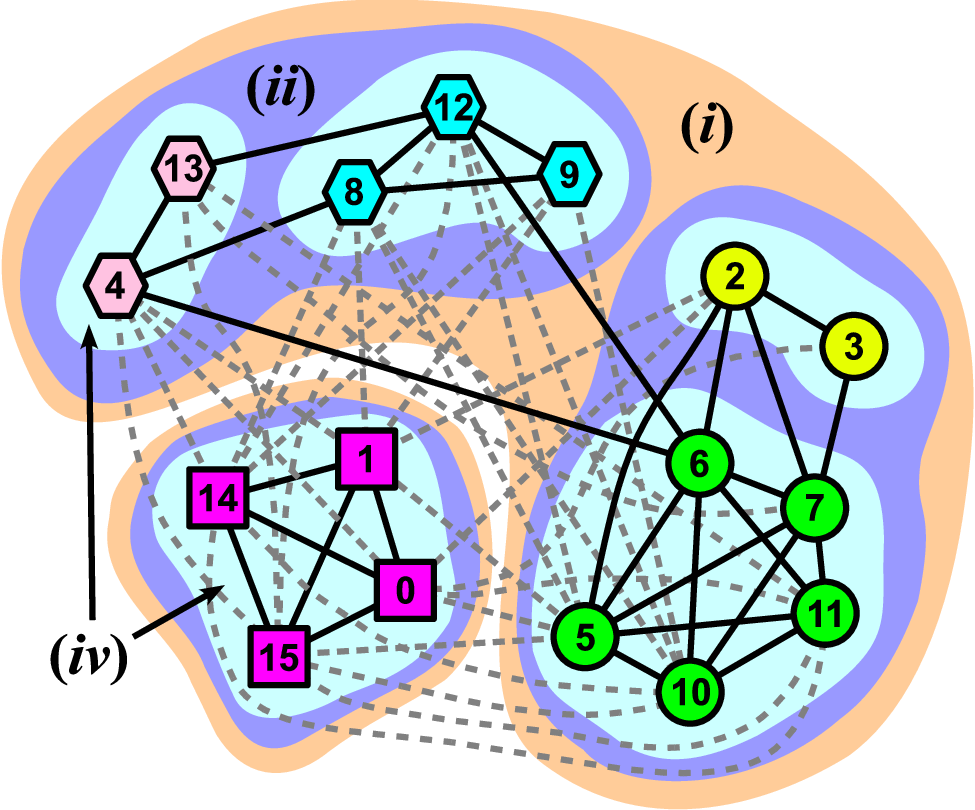}{
(Color online) 
Pictorial representation of $16$ Polopa tribes 
of Highland New Guinea~\cite{ref:highlandread,ref:highlandpage}.
Solid lines represent allied relationships, 
and gray dashed lines represent antagonistic relationships.
The three main levels of the structure are indicated
by shaded areas.
These groupings of tribes correspond to structures 
($i$), ($ii$), and ($iv$) in \figref{fig:highlandplot} 
in order of smaller group sizes.
Distinct node shapes (intermediate grouping) also 
correspond to structure ($ii$).
The three-cluster structure ($ii$) corresponds exactly 
to the analysis in~\cite{ref:highlandread,ref:highlandpage}.
Structure ($iii$) in \figref{fig:highlandplot} is formed when 
node $2$ joins the group at the bottom-right of the figure.
}{fig:highlandgraph}{3.0in}{t}

\myfig{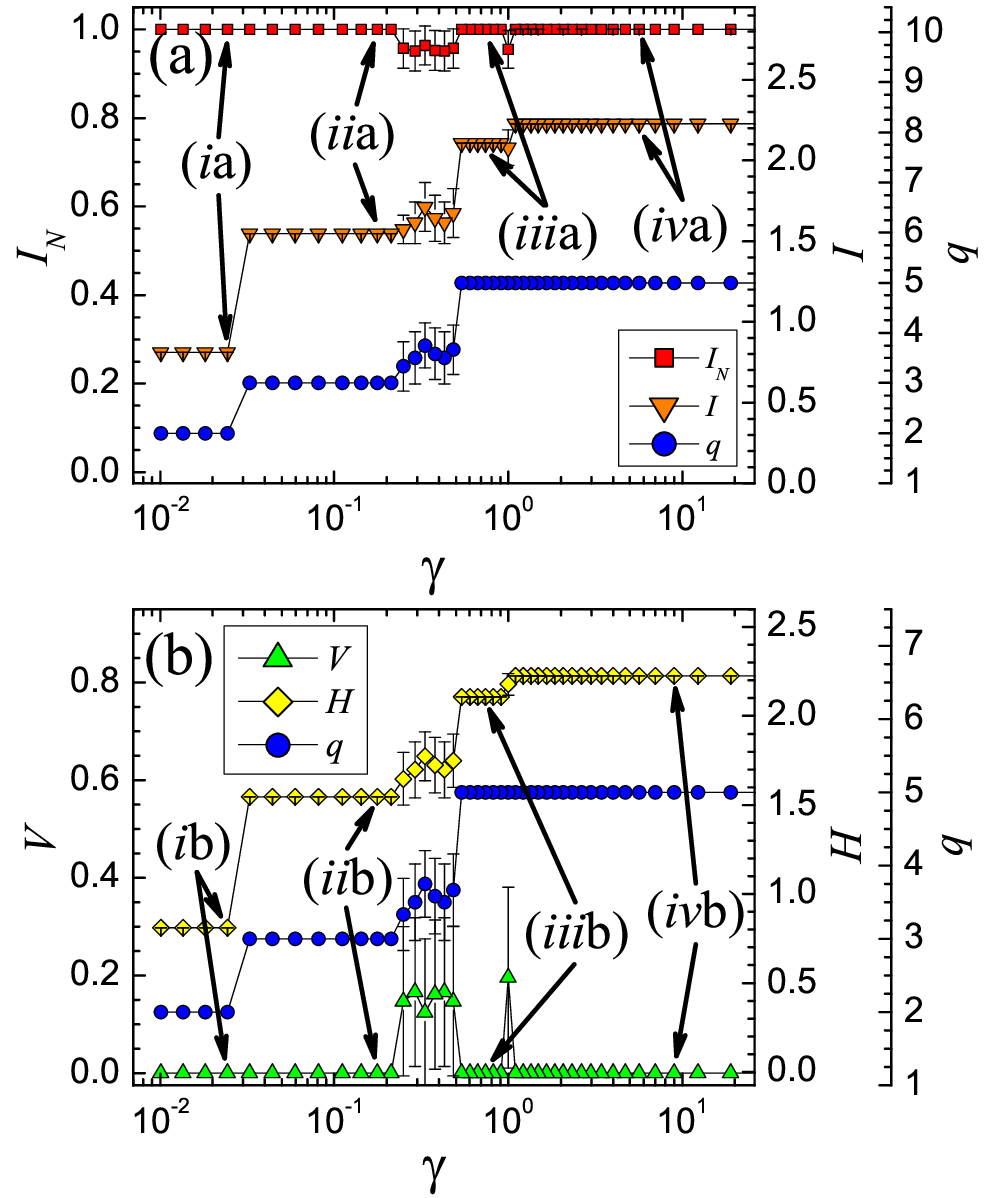}{
(Color online) 
Plot of information measures $I_N$, $V$, $H$, and $I$ 
in panels (a) and (b) vs. the Potts model weight 
$\gamma$ for $16$ Polopa tribes of Highland New Guinea.
The results are summarized in \figref{fig:highlandgraph}.
The right-offset axes plot the number of clusters $q$.
See \figref{fig:hierarchyplot} for a complete description
of the legends and axes.
The most important structure represented in the figure
is at ($ii$a,b) where the strong correlations agree 
exactly with data and analysis presented 
in~\cite{ref:highlandread,ref:highlandpage}.
See the text for a full discussion of the other
structures indicated in the figure.
}{fig:highlandplot}{3.0in}{t}

Figures \ref{fig:highlandgraph} and \ref{fig:highlandplot} 
%\figsref{fig:highlandgraph}{fig:highlandplot} 
show the results for $16$ Polopa tribes of Highland New 
Guinea~\cite{ref:highlandread,ref:highlandpage}.
These data feature allied, neutral, and antagonistic
relations between the sub-tribes of the region.
Hage and Harary \cite{ref:highlandpage} used symmetric edge 
weights of $+1$ for allied relations, $0$ for neutral relations, 
and $-1$ for antagonistic relations in their analysis; 
but these ``intuitive'' weight assignments are inconsistent 
if extended to systems that include few or no antagonistic 
relations
(such systems would tend to ``collapse'' into large groups).
Therefore, our model uses the more consistent assignments
of $-1$ for ``neutral'' relations 
and $-2$ for antagonistic relations.
Interestingly, Hage and Harary \cite{ref:highlandpage}
related the fact that the sub-tribes did not consider the
possibility of strictly neutral relations among tribes.
We use $12$ replicas with $10$ trials per replica to limit 
fluctuations in this very small data set
at a total run time of about $0.46$ sec~\cite{ref:computerused}.
We use an array data structure due to the missing edge weights.

Figure \ref{fig:highlandgraph} %\figref{fig:highlandgraph} 
depicts configurations
($i$), ($ii$), and ($iv$) from \figref{fig:highlandplot}
in order of smaller group sizes.
For presentation purposes, 
we allow three additional resolutions to be solved 
after the algorithm detects disjoint subgraphs at ($i$a,b).
Our three-cluster partition ($ii$) agrees exactly 
with those discussed in~\cite{ref:highlandpage}.
All configurations indicated in \figref{fig:highlandplot} 
are strongly defined with $I_N=1$ and $V=0$.
The first configuration ($i$) is a two-cluster solution 
which merges two sets of clusters of configuration ($ii$).
The small size of the system %(only $16$ nodes) 
causes the transition between configurations ($i$) 
and ($ii$) to be sharply defined.
To resolve the ambiguity, we must reference the plateaus
in the information measures $H$ or $I$ 
(or the number of clusters $q$~\cite{ref:multires}).

Strong NMI and VI values at ($iii$a,b) and ($iv$a,b)
correspond to two five-cluster solutions.
These solutions sub-divide the three-cluster system 
into two slightly different dense configurations 
of allied tribes. 
In configuration ($iii$), node $2$ is associated 
with the group on the bottom-right 
of \figref{fig:highlandgraph}.
In configuration ($iv$), 
all groups are cliques (maximally connected sub-graphs).
Both NMI and VI detect the transition between ($iii$) 
and ($iv$) with a short-lived spike.
The information measures $H$ and $I$ also show 
the transition with plateaus at different values.
Here, the number of clusters $q$ does not detect 
the transition since $q$ does not actually change.
Again, this is due to the limited variability in this system, 
but the same ambiguity occurs in \figref{fig:randomhierarchyplot}
for all three supplemental measures $H$, $I$, and $q$.

\section{Accuracy} \label{sec:accuracy}

% *** LFK Benchmark figures *****************************************

\myfig{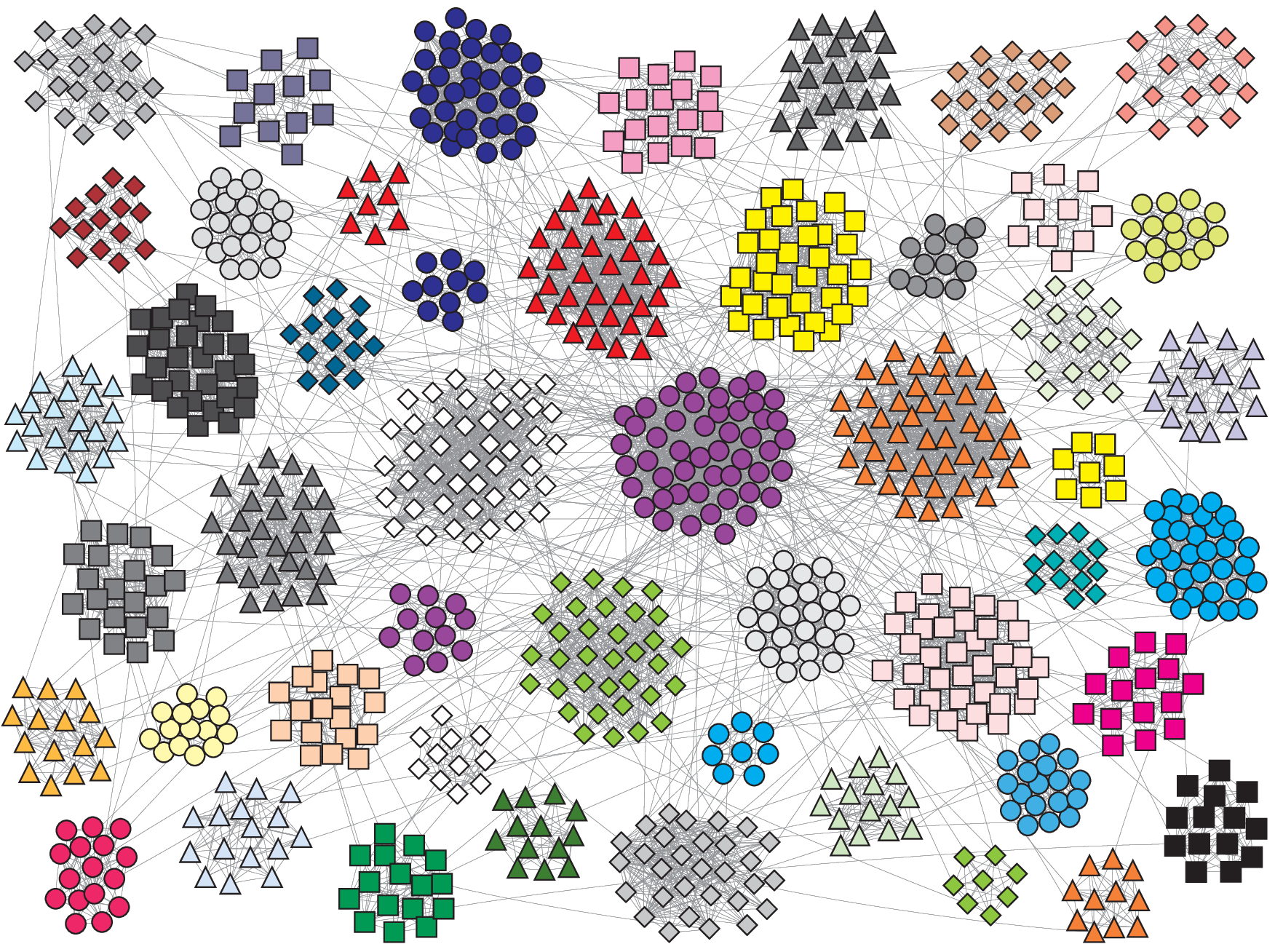}{(Color online) 
A sample graph with $N=1000$ nodes from the new benchmark 
proposed in~\cite{ref:lancbenchmark}. 
For presentation purposes, this depiction uses $\mu = 0.05$. 
Other parameters are $\alpha = 2$, $\beta = 1$,
$\langle k \rangle = 15$, and $k_{max} = 50$ (see text). 
}{fig:lfkgraph}{0.9\linewidth}{t}

\myfig{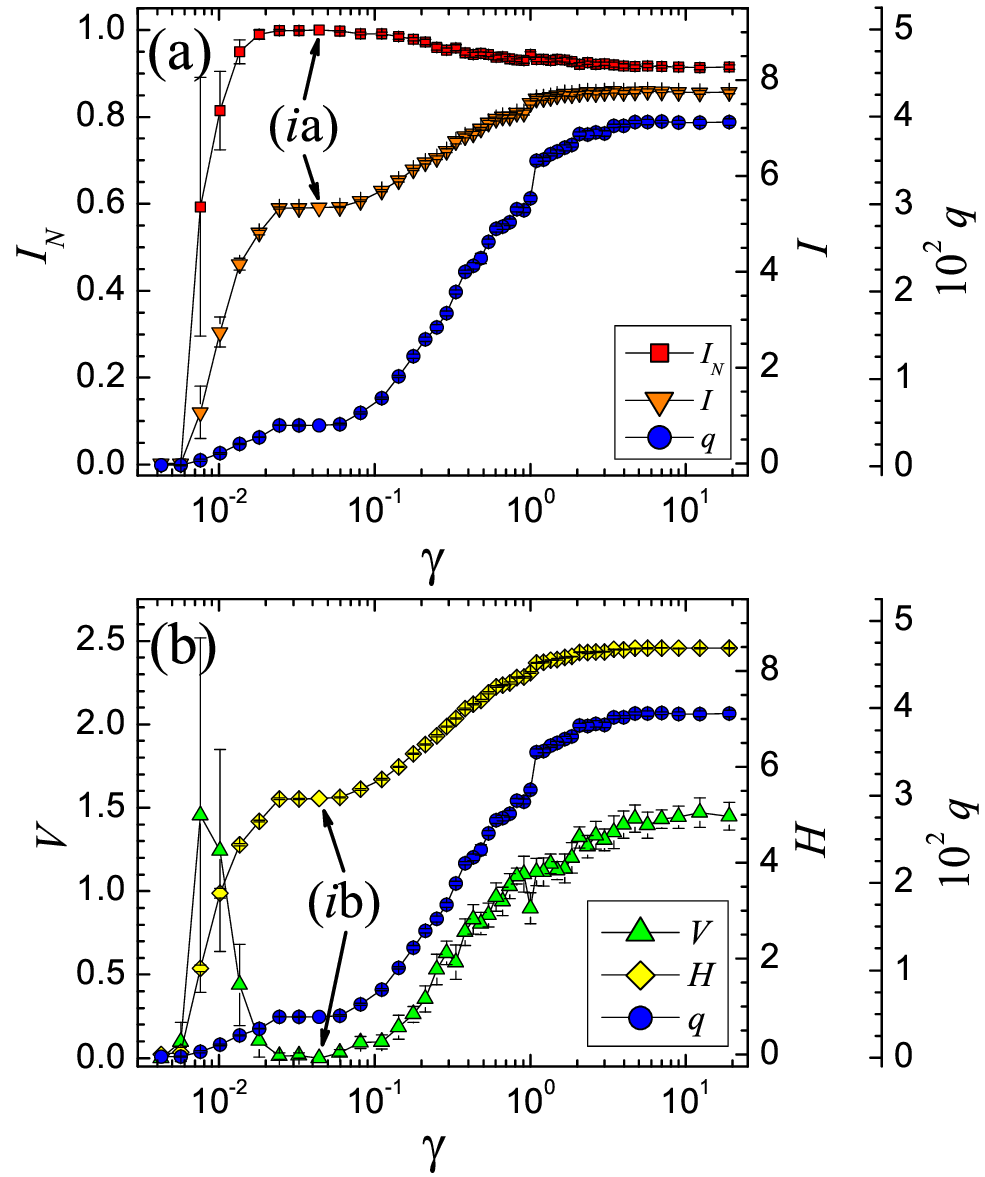}{(Color online) 
Plot of information measures $I_N$, $V$, $H$, and $I$ in panels 
(a) and (b) vs. the Potts model weight $\gamma$ 
for a single realization of the benchmark 
suggested in~\cite{ref:lancbenchmark}.
The right-offset axes plot the number of clusters $q$.
See \figref{fig:hierarchyplot} for a complete description
of the legends and axes.
Figure \ref{fig:lfkgraph} depicts a sample system from 
the benchmark (with a different mixing parameter $\mu$)
showing a distribution of community sizes.
This example plot is for $N = 1000$ at $\mu = 0.5$ 
where $50\%$ of each node's edges on average 
are connected to communities other than its own.
We use $\alpha = 2$ and $\beta = 1$ for the power-law 
distribution exponents of the node degrees 
and the community sizes respectively.
Using the algorithm in \secref{sec:algorithm},
we identify the strongest NMI and VI replica correlations 
among the different resolutions as the ``best" answer 
for the graph.
For this graph at $\mu = 0.5$,
there is only one extremal value of $I_N$ and $V$ which 
indicates that there is only one ``best" resolution 
for the defined system (see also Appendix \ref{app:lfkdetails}).
Note that these information values are the averages 
\emph{among the replicas}.
The full accuracy plot in \figref{fig:lfkaccuracyplot} 
plots the average $I_N$ between the ``\emph{best}'' 
partitions and the \emph{known} benchmark graphs
for a range of the mixing parameter $\mu$.
}{fig:lfksampleplot}{3.0in}{t}

\myfig{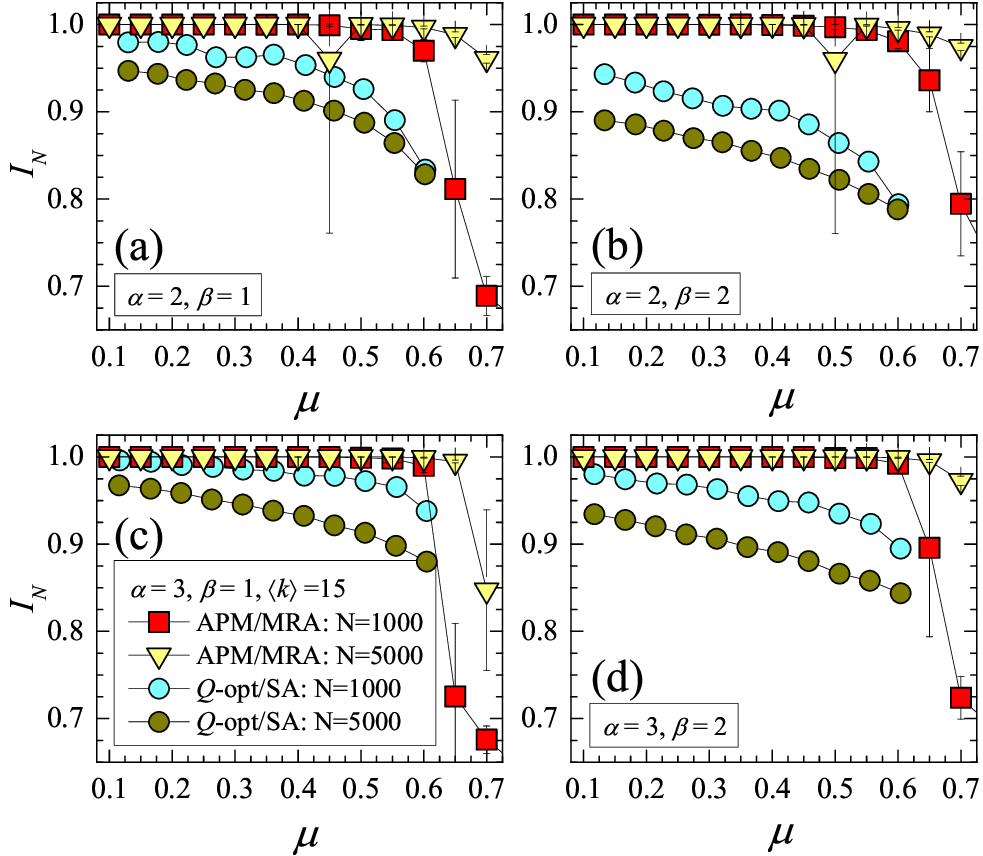}{(Color online) 
A plot of $I_N$ vs. $\mu$ for a new benchmark 
problem proposed in~\cite{ref:lancbenchmark}.
$I_N$ is calculated between the solved answer, by means 
the multiresolution algorithm in \secref{sec:algorithm}
using the absolute Potts model of \eqnref{eq:pottsmodel},
and the constructed benchmark graphs.
An example multiresolution analysis for one generated 
graph is in \figref{fig:lfksampleplot}.
$\mu$ is the fraction of edges of each node
(on average) that are assigned outside its own community.
We tested the power-law distribution exponents 
$\alpha = 2$ and $3$ and $\beta = 1$ and $2$
for the node degrees and the community sizes, respectively.
For comparison, we also plot the results 
from~\cite{ref:lancbenchmark} determined by modularity 
optimization ($Q$-opt) using simulated annealing. % (SA).
With the APM, our multiresolution algorithm %(MRA)
demonstrates extremely high accuracy for large systems 
(see text).  
Appendix \ref{app:lfkdetails} discusses the accuracy 
perturbations in panels (a) and (b) for $N=5000$ nodes.
Data for $N=1000$ and $N=5000$ nodes are averaged 
over $100$ and $25$ graphs respectively.
}{fig:lfkaccuracyplot}{0.9\linewidth}{t}

% *** end LFK Benchmark figures *************************************

% *** LFK Benchmark discussion **************************************
In \figdref{fig:lfkgraph}{fig:lfkaccuracyplot}, 
we test the accuracy of the multiresolution algorithm 
of \secref{sec:algorithm} with a recently proposed 
benchmark in~\cite{ref:lancbenchmark}.
An example graph with $N=1000$ nodes is depicted 
in \figref{fig:lfkgraph}.
This new benchmark can pose a significant challenge 
since it incorporates a more realistic heterogeneous 
%(power-law) 
distribution of community sizes and node degrees,
and it allows for testing across a large range 
of system sizes.
It divides a set of $N$ nodes into $q$ communities
with sizes assigned according to a power-law distribution 
with an exponent $\beta$.
The community sizes are optionally constrained by minimum 
and maximum sizes of $n_{min}$ and $n_{max}$.
The degrees of the nodes are also assigned in a 
power-law distribution with an exponent $\alpha$ 
with constraints specified by the maximum degree 
$k_{max}$ and the mean degree $\langle k \rangle$.
The minimum degree $k_{min}$ is set so that the 
distribution gives the correct mean $\langle k \rangle$.
A fraction $(1-\mu)$ of the edges of each node are 
connected to nodes within their own communities.
The remaining fraction $\mu$ are assigned to nodes
in other communities.

We test systems with $N=1000$ and $5000$ nodes and 
power-law exponents of $\alpha = 2$ and $3$ 
for the degree distribution and $\beta = 1$ and $2$ 
for the community size distribution.
We do not specify the optional community size 
constraints $n_{min}$ or $n_{max}$ allowing the benchmark
program to specify them by the degree distribution.
The node degree distribution is specified by 
$\langle k\rangle = 15$ and $k_{max} = 50$
where the mean degree $\langle k\rangle = 15$ was the 
most difficult of the tested values 
in~\cite{ref:lancbenchmark}.
We vary the mixing parameter $\mu$ in the range 
$0.1\le\mu\le 0.7$. 
The accuracy results are summarized 
in \figref{fig:lfkaccuracyplot}.

We apply the multiresolution algorithm of \secref{sec:algorithm} 
to identify the ``best" system partition.
Figure \ref{fig:lfksampleplot} %\figref{fig:lfksampleplot} 
shows an application of the 
algorithm for a single benchmark graph with $N=1000$, 
$\mu=0.5$, $\alpha=2$, and $\beta = 1$.
In this plot, we identify the ``best" system resolution
by the strongest average NMI correlation between all 
pairs of replicas.
We use $r=8$ replicas with $t=4$ energy optimization 
trials per replica.
As seen in \figref{fig:lfksampleplot}, both $I_N$ and $V$ 
(almost always) show only one extremal value which is 
the strongly defined system at ($i$a,b).
Plateaus in $H$, $I$, and $q$ qualitatively confirm
the structure indicated by the extrema in $I_N$ and $V$.
From these data, we determine that there is only one 
``best" resolution for the defined system.
See Appendix \ref{app:lfkdetails} for additional 
considerations in identifying the ``best" benchmark 
resolution.

In \figref{fig:lfkaccuracyplot}, we identify the ``best" 
partition for a set of benchmark graphs over a range 
of the mixing parameter $0.1\le\mu\le 0.7$. 
We then compare each solution via NMI with the 
``known'' partition.
We average over $100$ graphs for $N=1000$ and over $25$ 
graphs for $N=5000$ for each tested $\mu$.  
For comparison, we also include the results given 
in~\cite{ref:lancbenchmark} for modularity optimization %(Q-Opt) 
using a simulated annealing algorithm.
Combined with the APM %absolute Potts model (APM) 
of \eqnref{eq:pottsmodel},
our multiresolution algorithm performs excellently,
achieving almost perfect accuracy for each tested
distribution exponent $\alpha$ and $\beta$ 
and for a large range of the mixing parameter $\mu$.
The accuracy perturbations in panels (a) and (b) 
for $N=5000$ nodes are due to benchmark graphs 
with more than one local extremum in $I_N$ and $V$.
%(see Appendix \ref{app:lfkdetails}).
These perturbations are a result of the automated
selection of the single ``best" resolution
based on $I_N$ and $V$ extrema.  
We can largely eliminate them by a simple extension 
of the basic multiresolution 
algorithm (see Appendix \ref{app:lfkdetails}).
They are also nearly eliminated for these values of $N$ 
if we specify the default community size constraints 
of $n_{min}= 20$ and $n_{max}= 50$.

The absolute Potts model has little difficulty accurately 
solving the harder problem with $N=5000$ nodes because 
the edges connected to external communities are spread 
over more communities on average.
This construction causes a greater contrast of interior 
and external edge densities 
(considering edges connecting \emph{pairs} of communities).
This larger contrast allows the benchmark graph to be 
easily identified by the multiresolution algorithm. 
The converse occurs for small systems in the benchmark.

Our multiresolution algorithm has some difficulty 
in identifying all communities in this benchmark 
for exceptionally small systems ($N \lesssim 300$) 
where we achieve $I_N\simeq 1.0$ 
for a range of $\mu$ that increases with $N$
(for $N=300$, $I_N\simeq 1.0$ for $\mu\le 0.45$).
Communities are partitioned locally, independent 
of any \emph{global} parameters of the system; so
this limitation is not a resolution limit effect.
Rather, this behavior is due to simultaneously 
resolving communities with substantially 
different relative densities~\cite{ref:densitynote}.  
Palla \etal{}~\cite{ref:palla} stated that the community 
density should be used in identifying communities, 
which our Potts model does in effect.
In \secref{sec:model}, we suggested that it is the 
typical community edge density 
that characterizes the \emph{resolution} of a partition.
The difficulty in this benchmark is
due to defining communities by the fraction of each node's 
edges $(1-\mu)$ that lie within its own community.
Each community contains 
$\ell_s = n_s\langle k\rangle(1-\mu)/2$ edges on 
average where $n_s$ is the size of community $s$. 
The average edge density $p_s$ of community $s$ is 
%then $p_s = \langle k\rangle(1-\mu)/(n_s-1)$.
\begin{equation} \label{eq:lfkavgp}
  p_s = \frac{\langle k\rangle(1-\mu)}{(n_s-1)}.
\end{equation}
%The first two factors are constant on average 
The numerator is constant on average 
across all communities.
Our Potts model solves heterogeneously-sized systems well
(see \secsref{sec:smallhierarchy}{sec:largehierarchy}),
but one notable implication of \eqnref{eq:lfkavgp} is 
%the fact that very large communities can have 
%substantially lower community edge densities 
%compared to smaller communities in the system.
that the realistic distribution of community 
sizes leads to a substantial distribution of community 
edge densities with substantially different character 
for this benchmark.

Note also that our highly accurate results for 
$\mu = 0.6$ and $0.65$ for most values of $N$, $\alpha$, 
and $\beta$ in \figref{fig:lfkaccuracyplot}
show that the concept of a weak community 
structure~\cite{ref:radicchi},
where some nodes have more total edges connected 
to other communities than within their own,
is not too restrictive because the external edges 
can be dispersed among many other communities.
Indeed for $\mu>0.5$, all clusters in this 
benchmark on average exceed the definition of a weak
community since most, if not all, nodes have more 
exterior than internal edges.
So-called weak communities can occur frequently 
in social networks for example.
Individuals often know far more people 
than the size of the local ``community'' group(s) 
(friends, associates, etc.) of which they are members.
We showed a similar, but more striking, result when 
identifying level $3$ of the constructed hierarchy 
in Figs.\ \ref{fig:heterohierarchy}(a) 
and~\ref{fig:hierarchyplot} where the smallest 
communities had many more external than internal edges. 
Nevertheless, the model could easily resolve the 
communities at the correct resolution.

% *** begin Discussion discussion ***************************

\section{Discussion}
\label{sec:discussion}

In Figs.\ \ref{fig:hierarchyplot} -- \ref{fig:largehierarchyplot},
\ref{fig:dolphinsplot}, \ref{fig:highlandplot},
and \ref{fig:lfksampleplot},
strong correlations in NMI and VI appear to be consistent 
indicators of important multiresolution structures.
In most cases the assessments of the ``best" partitions
are confirmed by ``plateaus" in the mutual information $I$ 
and the Shannon entropy $H$.
These information plateaus are similar to those seen 
in the number of clusters $q$ in~\cite{ref:multires} and 
that are also observed in our data~\cite{ref:qaveragenote}.
In Ref.\ \cite{ref:multires}, the Arenas \etal{} indicated that plateaus 
in $q$ correspond to the most relevant system structures.
Our results largely affirm but also extend that observation.

In many pertinent applications of our algorithm, 
the final results (including, by fiat, our synthetic 
networks in \secsref{sec:smallhierarchy}{sec:largehierarchy})
are indeed hierarchical in the conventional sense.
That is, solving the Hamiltonian of \eqnref{eq:pottsmodel} 
anew with a different model weight $\gamma$ may break 
the communities apart, but it does not swap vertices 
between different communities at the correct resolutions.
As each resolution is solved independently
in our algorithm, we may (and indeed do) find more 
complicated multiresolution partitions where node 
reassignments lead to overlaps between communities that 
are perhaps disjoint on another level. 
This latter case is more subtle and appears
in systems such as the dolphin social network 
of \secref{sec:dolphins} and other individually 
oriented networks.

Variations in run time scaling among the different 
tests is influenced, sometimes strongly, by different 
levels of effective noise in each system
(aside from differing numbers of replicas and trials;
see Appendix \ref{app:cdtransition}).
For example, the hierarchy for \figref{fig:hierarchyplot} 
had a run time of $6.1$ s.
The corresponding random graph in \figref{fig:randomhierarchyplot},
with nearly the exact same density and number of nodes, 
finished in $6.9$ sec.

NMI and VI possess different strengths for quantitatively 
assessing multiresolution structure.
({\bf{1}})~Of course, NMI is normalized and VI is not
(although one normalization for VI is $1/\log_2 N$~\cite{ref:vi}).
Both of these features are useful.
({\bf{2}})~Figures \ref{fig:hierarchyplot}--\ref{fig:largehierarchyplot} 
show that VI more clearly identifies poor configurations.
In the high density regime ($\gamma\gtrsim 5$) 
of Figs.\ \ref{fig:hierarchyplot} and~\ref{fig:largehierarchyplot},
NMI shows a lower correlation compared 
to the peak values at ($i$) and ($ii$); 
but VI clearly indicates poor agreement.
In \figref{fig:randomhierarchyplot}, 
VI in panel (b) visually indicates a much poorer correlation 
in the $\gamma\simeq 0.3$ region as compared to NMI in panel (a).
({\bf{3}})~In \subfigref{fig:randomhierarchyplot}{a}, 
we identified peak ($i$a) as a ``trivial'' division with a huge 
component weakly connected to some small branch elements.
If one was actually interested identifying these very 
low-density solutions, NMI does identify them.
In panel (b), $V$ and $I$ simply indicate a 
very low-information configuration.

In many cases, extrema in either NMI or VI are sufficient 
to identify the multiresolution structure of a system.
Occasionally, we need to additionally reference the 
mutual information $I$ or the Shannon entropy $H$ 
(or the number of clusters $q$~\cite{ref:multires}). 
For example, in \figref{fig:hierarchyplot} NMI and VI almost 
do not distinguish between the $\gamma = 0.83$ partition 
(the exactly correct one) and the $\gamma = 1.6$ partition 
(one weakly connected node separates to form a new community)
because the separation between the two configurations
is almost imperceptible.
Both of these partitions correspond to level $3$ of the 
hierarchy depicted in \subfigref{fig:heterohierarchy}{a}, 
and both partitions have perfect correlations
($I_N=1$ and $V=0$).
In this case, the small changes in information measures 
$H$ and $I$ indicate a redundant $\gamma=1.6$ partition.
Also in \figsref{fig:lfksampleplot}{fig:lfkaccuracyplot},
we used the plateau to distinguish, when needed,
between strongly correlated transient partitions 
(due to random  elements of the benchmark generation process) 
and the more stable partition corresponding 
to the intended solution.

A similar challenge can occur for very small systems,
such as in the transition from ($i$) to ($ii$) 
in \figref{fig:highlandplot},
or for systems with few intercommunity connections.
As the resolution is adjusted in these systems, 
variability can be more limited; 
and system transitions can be sharply defined.
For these systems, it is possible that the NMI and VI 
correlations can remain strong and constant while crossing 
a structural transition.
In \figref{fig:highlandplot}, we avoid this ambiguity 
by noting that $H$ and $I$ clearly show a transition 
between structures ($i$) and ($ii$).
Such systems can also accentuate the perceived plateaus
in the multiresolution data because the variation 
in different configurations is small and transitions 
between major configurations can be sharp.
%Conversely, increasing levels of noise can mute 
%the appearance of the plateaus.

Given the distinctions, the two evaluations 
of multiresolution structure 
(``plateau'' behavior in $H$, $I$, and $q$ or strongly 
defined $I_N$ and $V$ correlations) are complimentary.
While the plateau behavior is important, %by itself 
it is a more qualitative assessment
of the ``best" resolutions for the system.
At least for our Potts model, 
under some conditions the plateaus in $H$, $I$, or $q$ 
can be weak enough to prevent them being used as the 
universal indicator of multiresolution structure.
In \figref{fig:randomhierarchyplot}, the plateaus even 
corresponded to a set of similarly sized partitions 
with similar densities rather than consistent structure.
The NMI and VI approach can more easily identify short-lived, 
but nevertheless strongly defined, structures 
(such as configuration ($iv$) in \figref{fig:dolphinsplot})
that the plateau criterion can miss.
In all Figs.\ \ref{fig:hierarchyplot} -- \ref{fig:largehierarchyplot},
\ref{fig:dolphinsplot}, \ref{fig:highlandplot}, 
and \ref{fig:lfksampleplot}, 
the major benefit of using the NMI and VI evaluations 
is that it appears to give a \emph{quantitative} estimate 
of the ``best" resolutions.
Together, the information measures appear to provide 
a consistent, accurate, and quantitative method of identifying
general multiresolution structure.

In further work, we will also consider a different method 
of adjusting the resolution of the system using the Hamiltonian
\begin{eqnarray}
	\Ham_{vt}(\{ \sigma \}) & = & 
	- \frac{1}{2} \sum_{i\neq j} \big[ 
	\left( a_{ij} + \alpha_{ij} \right) A_{ij} 
   - \left( b_{ij} + \beta_{ij}  \right) J_{ij} \big] \nonumber\\
	   & & %\phantom{-\frac{1}{2}\sum_{i\neq j}}
	   ~~~~~~~~~~\! 
	   \times\delta (\sigma_i,\sigma_j)	\label{eq:newH}
\end{eqnarray}
where $\alpha_{ij}$ and $\beta_{ij}$ are the new model 
weights as compared to $\gamma$ in \eqnref{eq:pottsmodel}.
This \emph{variable topology} Potts Hamiltonian 
is a generalized and continuous version 
of threshold cut-offs in weighted graphs.
It presents an alternative %, 
%and perhaps more ``natural" for some systems, 
method of continuously scaling the system by using 
an additive rather than a multiplicative scaling.
It differs from \eqnref{eq:pottsmodel} in that it progressively 
adjusts the topology of the system where multiplicative scaling 
does not change the system's connectedness.
Additive scaling may provide a different perspective 
on the evolution of the system structure over different 
scales, and it may better simulate how some real world 
models are ``stressed.''

Additionally, it may be possible to probe the system at a local 
level by using either localized partitions or by analyzing 
details within the confusion matrix at each resolution.
With this approach, we may be able to identify stable, 
but localized, structures beyond the information conveyed 
in the global information-based correlations.

We discovered and will report in detail in an upcoming
publication on a new sharp crossover between typical-easy 
and rare-hard community detection problems~\cite{ref:betadef}.
Our finding of a \emph{community detection transition}
constitutes an analog of the singular transition,
or more precisely, a singular region in the k-SAT 
(satisfiability) problem.
%M{\'e}zard, Parisi, and Zecchina~\cite{ref:mezardpz}
M{\'e}zard \etal{}~\cite{ref:mezardpz}
found that the hardest problems occur along well-defined
loci in the phase diagram of random satisfiability problems.
These loci of hard problems separate the SAT region 
(of satisfiable random problems) 
and the overly constrained UNSAT region
(in which the constraints cannot all be simultaneously satisfied).
We ascertained a similar phenomenon within community detection.
See Appendix \ref{app:cdtransition} for a summary of one facet 
of this transition.

Qualitatively, the analog of the SAT region is a common 
``easy'' and ``fast'' community detection region.
A ``transition" region, where computational cost rapidly 
increases and accuracy rapidly decreases, 
corresponds to the singular region of the k-SAT problem. 
A ``hard'' and ``slow'' community detection region corresponds 
to the UNSAT region of the k-SAT problem.  
For some community detection problems, 
the convergence rate can accelerate in the hard region due 
to the problem being rapidly trapped by local energy minima.

In a future work, we will detail the minimization 
of the ``free energy'' type functional 
of \eqnref{eq:replicainteraction}. 
This functional contains both the Potts model energy 
and the composite information function.
This latter information theory measure is maximized 
when the correlation between replicas is maximal.

\section{Conclusion}
\label{sec:conclusion}

%To conclude, 
We use a Potts model measure for community 
detection and apply it to detecting multiresolution structures:
({\bf{1}})~Our approach identifies and \emph{quantitatively} 
evaluates the `best' multiresolution structure(s), 
or lack thereof, in a graph.
({\bf{2}})~All resolutions are solved independently,
so the algorithm allows for the identification 
of completely general types of multiresolution structure.
({\bf{3}})~It is based on information comparisons, 
so in principle is should apply to any community detection 
model that can examine different resolutions.
({\bf{4}})~The underlying Potts model and algorithm are as 
accurate as the best methods currently available 
(see Appendix \ref{app:cdaccuracy}).
The model is a local measure of community 
structure, so it is free from the `resolution limit' 
as discussed in the 
literature~\cite{ref:rz,ref:fortunato,ref:kumpula,ref:multires,
ref:kumpulamultires,ref:resolutionlimitdef}. 
({\bf{5}})~Building on this foundation, 
the multiresolution algorithm demonstrates extremely 
high accuracy for large systems
using a recent benchmark proposed in~\cite{ref:lancbenchmark}
(see \secref{sec:accuracy}).
({\bf{6}})~We estimate that the computational cost scales 
as \bigO{N^{1+\beta}Z^{1+\beta}rt\log N\log Z} 
for some small $\beta$~\cite{ref:betadef,ref:logznote}
where  %$q$ is the number of communities,
$r$ is the number of replicas,
$t$ is the number of optimization trials per replica,
$Z$ is the average node degree, 
and $N$ is the number of nodes.
We have tested our community detection algorithm
on systems as large as \bigO{10^7} nodes and \bigO{10^9} edges 
(see Appendix \ref{app:cdlarge})~\cite{ref:computerused}.
The multiresolution algorithm requires a 
substantial number of individual community solutions; 
but due to the speed of the underlying algorithm, 
it can nevertheless examine systems over 
\bigO{10^5} nodes and \bigO{10^7} edges 
on a single-user workstation.
The algorithm should extend very efficiently to parallel 
or distributed computing methods allowing larger 
systems to be studied.

\section*{ACKNOWLEDGEMENTS}
%\begin{acknowledgments}
We thank UCINet and M. E. J. Newman for network data made 
available on their websites.
This work was supported by the LDRD DR on the physics 
of algorithms at LANL.
%\end{acknowledgments}

%\begin{appendices} % error *** not working correctly? ***
\appendix
\section{ACCURACY OF THE COMMUNITY DETECTION ALGORITHM}
\label{app:cdaccuracy}

\myfig{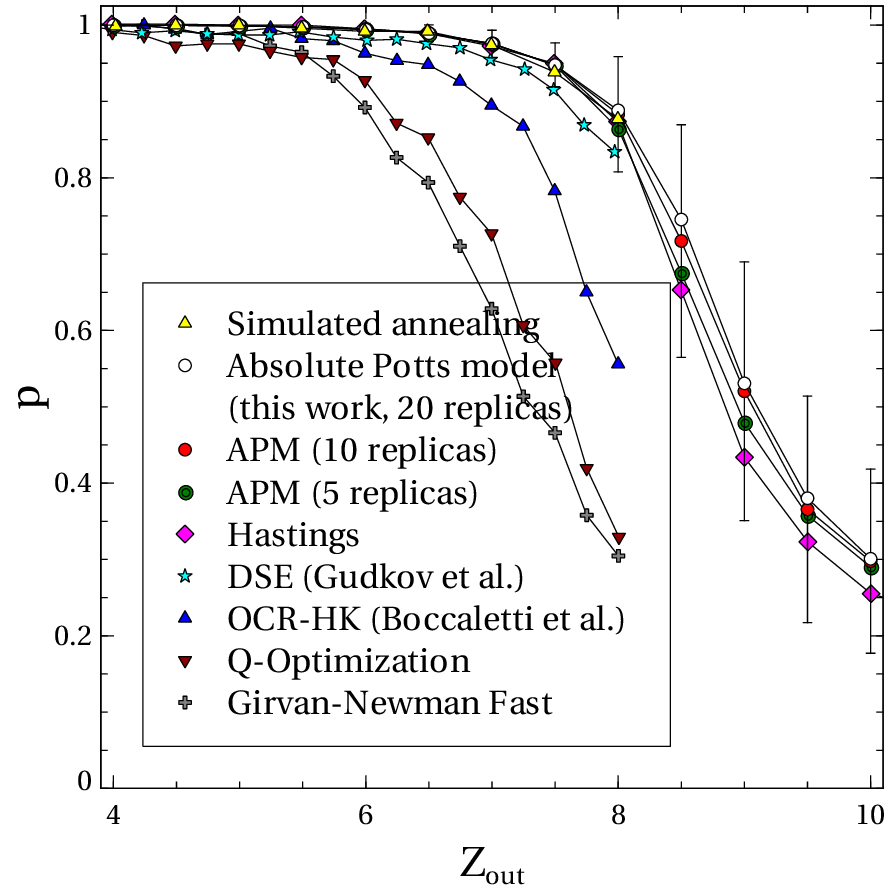}{(Color online) 
Reproduced from Ref.\ \cite{ref:rz}. 
A plot of the percentage of correctly identified nodes $p$ 
versus $Z_{out}$, the average number of edges 
that each node has connected to nodes outside of its own community.  
The average number of total edges per node is $Z=16$.
The benefit of extra trials $t$ reaches a point 
of diminishing returns around $t=10$ for many tests, 
and it is the intermediate difficulty trials 
($8\le Z_{out}\le 9$) that benefit the most 
from the additional optimization trials.
Note that the accuracy of our APM of \eqnref{eq:pottsmodel}
and algorithm in \secref{sec:cdalgorithm} is at least equal 
to the best algorithms.
Each point is averaged over $500$ systems.
}{fig:cdaccuracyplot}{0.8\linewidth}{t}

We demonstrate the accuracy of the community detection 
algorithm in \secref{sec:cdalgorithm} that is used 
to calculate the individual replica solutions 
in Step~\ref{step:solutions} 
of the multiresolution algorithm 
discussed in \secref{sec:algorithm}.
Our results using this frequent model problem in the 
literature were previously presented in~\cite{ref:rz}. 
The constructed model has $128$ nodes divided into $4$ 
clusters with $32$ nodes each.  
For each node, $Z_{in}$ edges are randomly 
connected to other nodes within its own community
and $Z_{out}$ edges are randomly connected to nodes 
in one of the other three communities.
The total degree of each node is $Z=Z_{in}+Z_{out}$
where we require an average degree of $Z=16$.

The task is to verify the defined community structure. 
In \figref{fig:cdaccuracyplot}, we use $\gamma =1$
in \eqnref{eq:pottsmodel} with $q$ constrained to four. 
We plot the percentage of correctly identified nodes $p$ 
versus the average number of externally connected edges 
per node $Z_{out}$.
We use the same measure of the ``percentage''
of correctly placed nodes as Ref.\ [$19$] 
within~\cite{ref:newmanfast}.
Four sets of data in \figref{fig:cdaccuracyplot} 
were assimilated by Boccaletti \etal{}~\cite{ref:boccaletti}. 
Simulated annealing~\cite{ref:danon} proved to be the most 
accurate algorithm in~\cite{ref:boccaletti} although 
it is computationally expensive.
Hastings~\cite{ref:hastings} and Gudkov \etal{}~\cite{ref:gudkov}
also demonstrated accurate results.

Our results in \figref{fig:cdaccuracyplot} use an older, 
slower version (without the neighbor-node search) 
of our algorithm.
For a small system of only $128$ nodes and $q=4$ by 
constraint, the difference in run time would be small.
For many of the tests, the benefit of extra trials $t$ 
reaches a point of diminishing returns by $t=10$. 
High noise systems rapidly trap different replicas 
in local energy minima, so it is the ``intermediate'' 
difficulty solutions ($8\le Z_{out}\le 9$) that benefit 
the most from additional optimization trials. 
Our method maintains an accuracy rate at least equal
to the best available algorithms.
In particular, it maintains a $95\%$ or better accuracy 
rate up to $Z_{out}=7.5$.

\section{TRANSITION EFFECTS OF NOISE LEVEL 
ON COMMUNITY DETECTION ACCURACY}  
\label{app:cdtransition}

\myfig{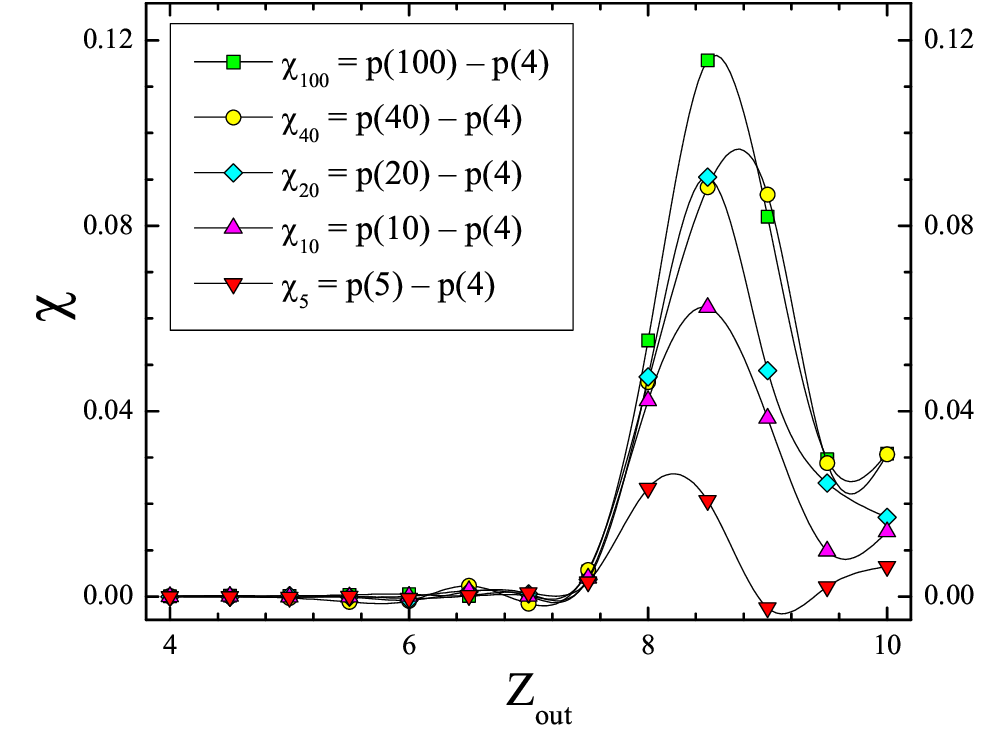}{(Color online) 
A plot of the susceptibility  $\chi_n \equiv p(t=n) - p(t=4)$ 
versus $Z_{out}$, the average number of edges 
that each node has connected exterior to its own community.
$\chi_n$ is the percentage increase in the accuracy 
of each test as the number of trials $t=n$ is increased 
from $n=5$ to $n=100$.
The average number of total edges per node is $Z=16$.
$p$ is the percentage of correctly identified
nodes from \figref{fig:cdaccuracyplot}.
%, and $t$ is the number of trials at each test.
The curves are spline fits and are intended 
for visualization purposes only.
Additional trials are unnecessary in the easy
region $Z_{out}\lesssim 7$.
The benefit of extra trials is largest in the short 
transition region $8\le Z_{out}\le 9$.
Afterwards, the benefit diminishes into the hard region 
$Z_{out}\gtrsim 9.5$ where the accuracy improvement is small 
even with a large number of attempted optimization trials.
}{fig:chiplot}{0.9\linewidth}{t}

The benchmark problem that serves as the basis 
for data in \figref{fig:chiplot} is discussed in detail 
in Appendix \ref{app:cdaccuracy}.
In \figref{fig:chiplot}, we plot for several numbers 
of trials $n$,
the ``susceptibility'' $\chi_{n} \equiv p(t=n) - p(t=4)$ 
versus $Z_{out}$, the average number of edges 
that each node has connected exterior to its own community.
%$t$ is the number of optimization trials attempted.
The average number of total edges per node is $Z=16$.
$p$ is the percentage of correctly identified nodes 
from \figref{fig:cdaccuracyplot} 
(see Ref. [19] in~\cite{ref:newmanfast}),
and $t$ is the number of trials at each test.
The ordinate $\chi$ in \figref{fig:chiplot} is the 
percentage improvement in accuracy based on the number 
of optimization trials that are used.

As $Z_{out}$ increases, the noise in the system increases.
Figure \ref{fig:chiplot} illustrates how the noise 
in the system affects the effort required to solve 
the system as accurately as possible. 
The benefit of extra optimization trials is negligible 
for the easy region up until about $Z=7$. 
Additional trials become more important 
for a short transition region ($8\le Z_{out}\le 9$).
Afterwards, the benefit of additional trials quickly 
reaches a point of diminishing returns in the hard
region $Z_{out}\gtrsim 9.5$ where it fails 
to produce large improvements in accuracy despite 
significantly more computational effort. 

As the number of trials $n$ increases, the ``susceptibility'' 
$\chi_{n}$ progressively exhibits a more pronounced peak.
Such a trend is also evidenced in the susceptibility 
of finite size physical systems.  
We have also identified a similar and related dynamic 
feature of the transition that is quantified by the increased 
computational time required for a single solution~\cite{ref:betadef}
(beyond any added computational cost due to extra 
energy optimization trials).

\section{COMMUNITY DETECTION OF A LARGE SYSTEM}
\label{app:cdlarge}

We tested our community detection algorithm in 
\secref{sec:cdalgorithm} using the neighbor-node 
search on a synthetic network with over one billion links.
We generated a random set of $N=40$ million nodes separated 
into $1.25$ million heterogeneously-sized communities 
with sizes ranging from $10$ to $62$ nodes.
(Note that it is the number of edges that limits the calculation
as opposed strictly to the number of nodes.)
The random edge connection probability for the communities 
was $p_{in}=0.9$.
Nodes between these communities were connected with a probability
of $p_{out}=5.31\times 10^{-7}$.
Each node has an average number of interior and exterior 
edges of $Z_{in}\simeq 29$ and $Z_{out}\simeq 21$.
The total number of edges was $1~000~211~862$.
The average density of the graph was $p=1.25\times 10^{-6}$.

We used $\gamma =1$ in \eqnref{eq:pottsmodel} and the 
algorithm in \secref{sec:cdalgorithm} to solve the system.
There were $13$ nodes that were not placed within their 
intended communities.  
These are likely due to random initialization fluctuations.
The information correlations for the ``known'' and solved 
answers were %$I_N>0.999~\!999$ 
$I_N\simeq 1.00$ and $V=1.85\times 10^{-6}$ 
with $V_{max}=\log_2 N\simeq 25.3$.
Both of these measures indicate very strong agreement.
The total calculation time was $3.7$ h 
on a single processor~\cite{ref:computerused}.

%\newpage \

%\newpage

\section{GENERALIZATION OF THE INFORMATION-BASED REPLICA METHOD}
\label{app:generalentropy}

In \secref{sec:algorithm}, 
we may recast the information theory measures used
to evaluate the correlation between different replicas 
for other (non-graph theoretic) optimization problems 
with general Hamiltonians (or cost functions) $\cal{H}$.
An alternate form of \eqnref{eq:mi} for the mutual 
information between replicas $i$ and $j$ is
\begin{equation}  \label{eq:generalH}
  I(i,j) = H(i) + H(j) - H(i,j)
\end{equation}
%where $H(i)$, $H(j)$, and $H(i,j)$ denote, respectively, 
%the entropy of replica $i$, the entropy of replica $j$, 
%and the entropy of the combined system formed by replicas 
%$i$ and $j$. 
where $H(i)$, $H(j)$, and $H(i,j)$ denote 
the entropies of replica $i$, replica $j$, and the combined 
system formed by both replicas, respectively. 
%Instead of using \eqnref{eq:mi} for the mutual information,
Instead of using \eqnref{eq:mi}, %for the mutual information,
we write the Shannon entropy $H(i,j)$ for the combined 
replicas $i$ and $j$ which we then apply in \eqnref{eq:generalH}.
For general Hamiltonians ${\cal{H}}$, we replace $H(i)$, $H(j)$, 
and $H(i,j)$ by a thermodynamic entropy for the respective systems.
%(see Appendix \ref{app:generalentropy}).
%\cite{ref:MInote}.

In the general case, the thermodynamic entropy $H(i,j)$ 
of the system formed by the union of replicas $i$ and $j$ is
%and the entropy $H(i)$ of system $i$ are given respectively by 
%\begin{eqnarray} 
%H(i,j) &=& \frac{\partial}{\partial T} 
%  \Bigg\{ \beta^{-1} \log \bigg[ \mathrm{Tr}_{i,j} \Big(
%     e^{-\beta {\cal{H}}(i)} + e^{-\beta{\cal{H}}(j)}
%  \Big) \bigg] \Bigg\}, \nonumber\\ 
%H(i) &=&  \frac{\partial}{\partial T} 
%  \Bigg\{ \beta^{-1} \log \bigg[ \mathrm{Tr}_{i} \Big(
%     e^{-\beta {\cal{H}}(i)} 
%  \Big) \bigg] \Bigg\},
%  \label{eq:hab}
%\end{eqnarray}
\begin{equation} 
H(i,j) = \frac{\partial}{\partial T} 
  \bigg\{ \beta^{-1} \log \bigg[ \mathrm{Tr}_{i,j} \Big(
     e^{-\beta {\cal{H}}(i)} + e^{-\beta{\cal{H}}(j)}
  \Big) \bigg] \bigg\},
  \label{eq:hab}
\end{equation} 
and the entropy $H(i)$ of system $i$ or $j$ is
\begin{equation} 
H(i) = \frac{\partial}{\partial T} 
  \bigg\{ \beta^{-1} \log \bigg[ \mathrm{Tr}_{i} \Big(
     e^{-\beta {\cal{H}}(i)} 
  \Big) \bigg] \bigg\}.
  \label{eq:ha}
\end{equation}
%where ${\cal{H}}(i)$ is the Hamiltonian of replica $i$, 
%${\cal{H}}(j)$ is the Hamiltonian of replica $j$, 
${\cal{H}}(i)$ and ${\cal{H}}(j)$ are the 
Hamiltonians of replicas $i$ and $j$,
and $\beta=1/(T\ln 2)$ is the inverse temperature.
Within our approach, an ensemble reduces to a finite 
number of points (replicas) whose correlations 
are monitored by information theory measures.
%\eqnarefs{eq:hab}{eq:ha} are the entropies
%for the system composed of replicas $i$ and $j$.
This form pertains to the general case in which both 
$i$ and $j$ pertain to a collection of decoupled %multiple 
copies, and the traces are over all coordinates 
in replicas $i$ and $j$. 
%We likewise use the corresponding thermodynamic relations 
%for $H(i)$ and $H(j)$ by keeping only one term in the trace 
%for each.

The standard mutual information of \eqnref{eq:mi} 
is generally not invariant (as it ideally should be) 
under the permutation of ``identical" nodes
(those with an identical neighbor list that are otherwise 
indistinguishable by other parameters of the system).
Specifically, we refer to nodes $i$ and $j$ as identical
in a graph %, defined by an adjacency matrix $A$, 
if the adjacency matrix $A$ is invariant under the 
permutation of node $i$ with node $j$~\cite{ref:entropydifferent}.
%The invariance in the mutual information can also be 
%ameliorated by a more general definition of the confusion 
%matrix as we will discuss elsewhere. 
%since every node is implicitly assumed to have a unique identity.
That is, $A$ commutes with the permutation of nodes $i$ and $j$,
$[P_{ij},A]=0$, if nodes $i$ and $j$ are identical.
The thermodynamic entropies of \eqnarefs{eq:hab}{eq:ha} are invariant 
under permutations of identical nodes because any symmetries, 
or lack thereof, are fully represented in the system 
Hamiltonian $\cal{H}$.

In the simplest case with only one copy of the 
system in replica $i$ and one copy in replica $j$,
there is only one term in both $i$ and $j$;
and the designation $\mathrm{Tr}_{i,j}$ becomes 
redundant (the entropies of $i$ and $j$ are also 
trivially $H(i) = H(j) =0$).
In a more realistic approximation to thermodynamic 
quantities, each of the replicas $i$ and $j$ contain 
a number of independent decoupled copies of the system.
Inserting \eqnarefs{eq:hab}{eq:ha}
%and the associated relations for $H(i)$ and $H(j)$ 
into \eqnref{eq:generalH}, 
we obtain the mutual information between $i$ and $j$. 
%The normalized mutual information is then given
%by \eqnref{eq:nmi}. 
NMI and VI are then given by \eqnarefs{eq:nmi}{eq:vi}. 
Other information measures $S(i,j)$ 
between replicas $i$ and $j$ may also be computed. 
Along similar lines, multi-replica (higher than two) 
forms may replace the sum over two-replica configurations 
in \eqnarefs{eq:replicainteraction}{eq:hab}.
%in Eqs.\ (\ref{eq:replicainteraction}), (\ref{eq:hab}), 
%and (\ref{eq:ha}).

We may also reconstruct the information measures %,
%by an effective Hamiltonian, 
using a different physical analogy. 
The Shannon entropy of \eqnref{eq:shannonentropy} 
is analogous to an ensemble where each of the $N$ 
nodes corresponds to one point in the ensemble.
The communities correspond to $q$ possible states 
of a single particle with energies $\{E_{k}\}$ 
for $k=1$ to $q$ at a given temperature $T$ such 
that the same community occupation probabilities 
are reproduced as
%\begin{equation}
 %p_k=\frac{e^{-\beta E_k}}{\sum_{i=1}^q e^{-\beta E_i}}
%\end{equation}
$p_k=n_k/N=e^{-\beta E_k}/\sum_{i=1}^q e^{-\beta E_i}$
where the inverse temperature is $\beta = 1/(T \ln 2)$. 
The mutual information $I$ of \eqnref{eq:mi} is equivalent 
to an ensemble of size $N$ for a two-particle system 
in which each particle can be in any of $q$ states. 
The interaction between the two particles is such 
that it leads to energies $\{E_{ij}\}$ for the two 
occupied communities $i$ and $j$.
These interactions lead to a relative probability 
$p_{ij} = n_{ij}/N$ for occupying the two-particle 
states that is proportional to $e^{-\beta E_{ij}}$.
%$p_{ij}=e^{-\beta E_ij}/\sum_{i,j=1}^q e^{-\beta E_i}$.
%where the inverse temperature $\beta = 1/(T \ln 2)$. 
The effective Hamiltonian for the resulting physical 
system does not directly depend on the identities 
of the $N$ nodes (although it does not distinguish 
between ``identical" and distinguishable nodes).

One potential limitation of our thermodynamic framework 
in \eqnarefs{eq:hab}{eq:ha} is that general, 
non-graph theoretic, applications may require many copies 
of the same system. 
The traces $\mathrm{Tr}_{i}$, $\mathrm{Tr}_{j}$
need to be calculated on multiple copies of the same system. 
This is bypassed in the application of mutual information 
for graph problems because the node number $N$ effectively plays the 
role of many ensemble points (multiple replica copies) 
on which the thermodynamic average is to be taken.

\section{MULTIRESOLUTION BENCHMARK COMMENTS}
\label{app:lfkdetails}

% *** LFK dual maximum plot **********************************
\myfig{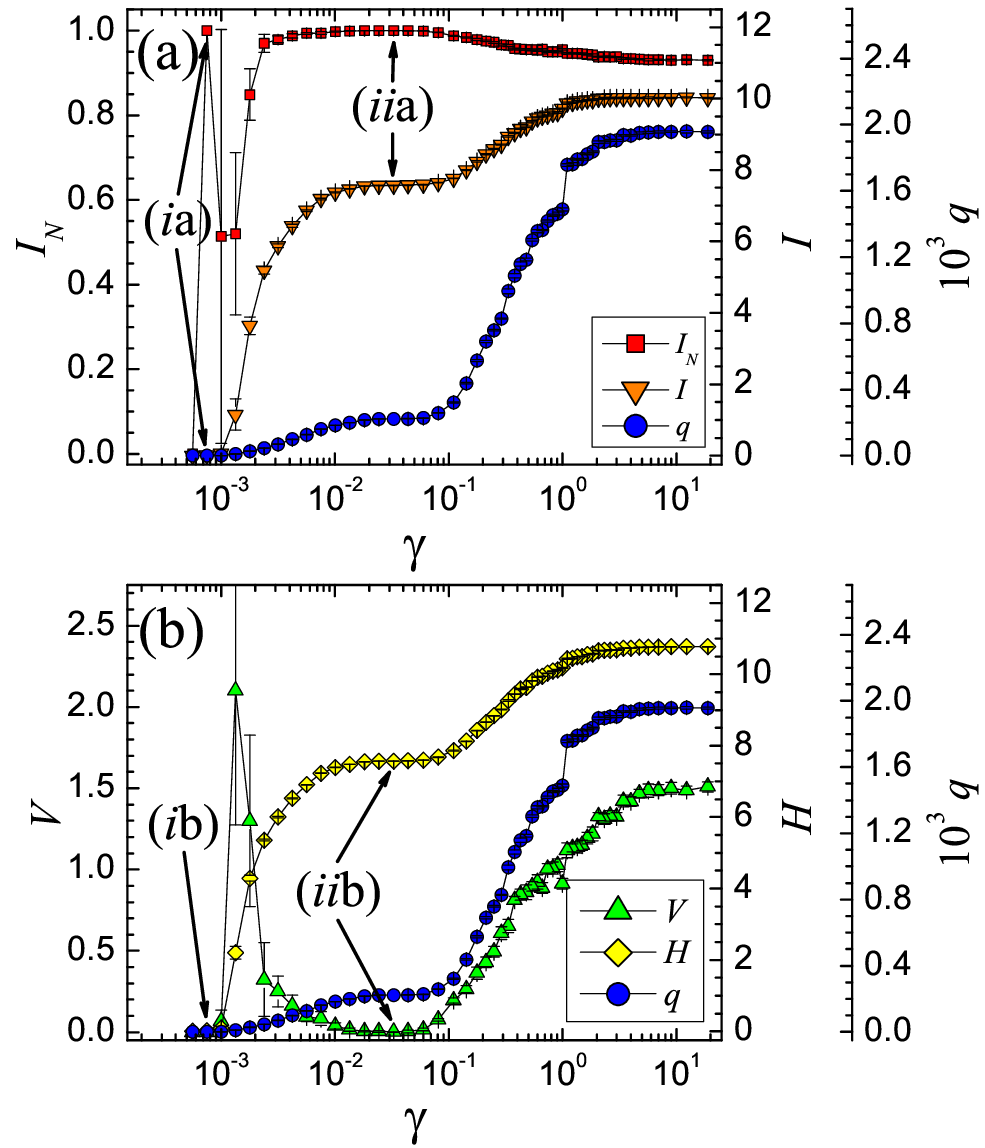}{(Color online) 
Plot of information measures $I_N$, $V$, $H$, and $I$ 
in panels (a) and (b) vs the Potts model weight 
$\gamma$ for a single realization of the benchmark 
suggested in~\cite{ref:lancbenchmark}.
The right-offset axes plot the number of clusters $q$.
See \figref{fig:hierarchyplot} for a complete description
of the legends and axes and \secref{sec:accuracy}
for a full explanation of the benchmark.
This example plot is a multiresolution analysis 
for $N = 5000$ at $\mu = 0.45$ where $55\%$ of each 
node's edges on average are connected to communities 
other than its own.
The power-law distribution exponents for the node 
degrees and the community sizes are $\alpha = 2$ 
and $\beta = 1$, respectively.
We use the algorithm in \secref{sec:algorithm} to attempt 
to identify the ``best" resolution for the graph.
For some cases in the benchmark, 
such as this graph at $\mu= 0.45$,
there exists more than one extremal value of $I_N$ and $V$
where the low-density configuration at ($i$a,b) also has 
slightly stronger NMI and VI correlations 
($\delta I_N\simeq 6.3\times 10^{-5}$) 
than the intended benchmark answer at ($ii$a,b).
In this example case, a casual inspection indicates that 
the stable region at ($ii$a,b) is clearly the ``best" 
partition which also corresponds almost exactly 
to the intended benchmark solution.
The automated version of the algorithm favors the slightly 
stronger %(in terms of NMI and VI correlations) 
low-density configuration at ($i$a,b) 
as the ``best" resolution for the graph.
}{fig:lfkproblemplot}{3.0in}{t}

% *** end of LFK dual maximum plot **************************

As discussed in \secref{sec:accuracy}, we used the new 
benchmark problem presented in~\cite{ref:lancbenchmark} 
to test the accuracy of our multiresolution algorithm
of \secref{sec:algorithm}.
Our algorithm attempts to identify 
all strongly defined resolutions.
By design, the benchmark in~\cite{ref:lancbenchmark} 
constructs a ``realistic'' system with a single intended 
solution; however, random effects of the graph generation 
process can also create additional transient, 
but nevertheless strongly defined resolutions which our 
algorithm can detect.
In implementing the benchmark, we endeavor to automate 
the identification process to determine the single 
``best" resolution as intended by the benchmark. 
We explain two special cases.
%for identifying the ``best" resolution.

The first difficulty is encountered for $\mu\lesssim 0.4$.
We can detect multiple resolutions with perfect
correlations ($I_N = 1$ and $V = 0$) 
for resolutions near the intended benchmark solution 
which occur more frequently as $\mu$ decreases.
This effect is similar to partition ($i$) that occurred 
near partition ($ii$) in \figref{fig:highlandplot}.
The transitional resolutions are not necessarily meaningless
partitions on an individual graph-by-graph basis,
but they are artifacts of the randomly generated system 
and thus vary across the different benchmark graphs.
Similar to structure ($ii$a,b) in \figref{fig:lfkproblemplot}, 
the plateaus in the information measures $H$ or $I$ 
(or the number of clusters $q$~\cite{ref:multires}) 
indicate a more ``stable'' partition.
It is this stable partition that corresponds to the 
intended solution for the benchmark graph.
Thus, when necessary, we use the plateaus to discriminate 
between the short-lived and the most stable strongly 
defined partitions in order to determine the single ``best" 
resolution for each benchmark graph.

A second difficulty %of automating the multiresolution 
is shown in \figref{fig:lfkproblemplot} which occurs 
most frequently in the range of mixing parameter
$0.45\lesssim\mu\lesssim 0.65$.  
The stable configuration that corresponds to the 
intended benchmark answer is configuration ($ii$a,b).
The low-density, transient, but strongly correlated 
configuration at ($i$a,b) has a slightly higher NMI 
correlation.
Even a casual visual inspection of the data 
in \figref{fig:lfkproblemplot}
indicates that configuration ($ii$a,b) is 
the dominant configuration for the graph.
Specifically, configuration ($ii$a,b) possesses 
both very strong NMI and VI correlations 
($I_N\simeq 1.0$ and $V\simeq 0.0$)
as well as stable and long $H$, $I$, and $q$ plateaus,
and indeed it corresponds almost exactly 
to the intended benchmark answer.
However, the automated application of the multiresolution 
algorithm slightly favors configuration ($i$a,b) 
as the ``best" resolution since it has a higher NMI 
($\delta I_N\simeq 6.3\times 10^{-5}$) and a lower VI.
(See \secsref{sec:randomgraph}{sec:discussion} regarding 
potential problems of using the plateaus in $H$, $I$, or $q$ 
as the primary measure for identifying the ``best" resolutions.)

These graphs are the cause of the accuracy perturbations 
in Figs. \ref{fig:lfkaccuracyplot}(a) and \ref{fig:lfkaccuracyplot}(b).
They are less frequent for $\beta =2$ since the community 
size distribution is more skewed towards smaller communities
than for $\beta = 1$.
We note that the average accuracy for the perturbations 
in Figs. \ref{fig:lfkaccuracyplot}(a) and \ref{fig:lfkaccuracyplot}(b) 
is still high at $I_N\simeq 0.96$.
In \figref{fig:lfkaccuracyplot}, an iteration cap acted 
as an effective filter for most low-density spikes. 
We could further improve the automated analyses of such 
graphs by replacing this filter with moving NMI or VI 
averages (\ie{}, each moving average is over the 
NMI or VI of several nearby resolutions) to exclude
resolutions such as the short-lived configuration ($i$).
%Then the extremal NMI or VI values and plateau criteria 
%will even more accurately identify the most strongly defined 
%and stable systems. 

%there is an error with \begin{appendices}... \end{appendices}
%\end{appendices} 

% ********************** END DOCUMENT TEXT **************************
%\bibliography{hierarchypaper}  % hierarchypaper.bib is the name of our database

\begin{thebibliography}{57}
\bibitem[{ref({\natexlab{a}})}]{ref:multiresusage}
\bibinfo{note}{We adopt the term ``multiresolution'' as used
  in~\cite{ref:kumpulamultires} to indicate that this algorithm is not limited
  to hierarchical structures.}
\bibitem[{ref({\natexlab{b}})}]{ref:overlapdef}
\bibinfo{note}{Overlapping nodes are those that are shared between multiple
  communities.}
\bibitem[{ref({\natexlab{d}})}]{ref:computerused}
\bibinfo{note}{Graphs were solved on a single core 
of an Intel Core 2 Quad Q6600 at $2.4$ GHz with $2$ GB random 
access memory (RAM) or on Opteron systems at $2.2$ -- $2.8$ GHz 
with $4$ -- $32$ GB of RAM.}
\bibitem[{ref({\natexlab{e}})}]{ref:clustermergenote}
\bibinfo{note}{Additional work during each replica solution (Step
  \ref{step:solutions} of the multiresolution algorithm) includes 
  two optimizations. 
  First, as in~\cite{ref:rz} each replica can utilize multiple
  trials $t$ to mitigate the effects of local energy minima 
  by selecting the lowest energy trial for that replica. 
  Typically $t\simeq 4$, but it can be higher $t\simeq 20$ 
  for difficult systems.
  Second, the algorithm manually merges clusters that will lower 
  the energy of the system. 
  This effect arises because the algorithm evolves the system
  by sequentially moving one node at a time to a new community 
  based on the largest energy decrease for the node given the 
  current state of the system.
  Some edge configurations (particularly heavily weighted graphs 
  with $\gamma\ll 1$) can hinder the process of merging clusters 
  when moving one node at a time.
  The number of merges is indefinite, but it is generally small 
  enough to not significantly alter the overall performance 
  of the algorithm.
  In some cases, additional node-level adjustments may follow 
  as a result of any manual merges. 
  The computational cost is estimated to be \bigO{NZ\log Z\log q} 
  except for $\gamma\ll 1$, where $q$ is small.}
\bibitem[{ref({\natexlab{f}})}]{ref:betadef}
\bibinfo{note}{The scaling dependence of $\beta$ is a complicated function
  $\beta\equiv\beta(N,q,Z_{in},Z_{out},\gamma)$, where $N$ is the number of
  nodes, $q$ is the number of communities, $\gamma$ is the Potts model weight,
  and $Z_{in}$ and $Z_{out}$ are the average number of interior and exterior
  connected edges for each node with respect to its own community. A typical
  average scaling is $\beta\lesssim 0.3$. Empirically for large graphs, worst
  case behavior is seen in localized regions with an intermediate to high
  $Z_{out}/Z_{in}$ ratio where there is significant system confusion (where any
  algorithm would have difficulty). In these cases $\beta\simeq 9$, and the
  system is in transition to a structure that is more difficult to solve. The
  transition effect is localized and is not representative of the global
  scaling. Afterward, $\beta$ drops to a value more representative of the
  average scaling. As the system becomes increasingly difficult, bordering on
  incoherent, the convergence rate can actually accelerate as the algorithm is
  rapidly trapped by local minima. The rapid transition is representative of
  general transitions between typical-easy and rare-hard problems and
  constitutes an analog of the singular transition point the k-SAT problem. We
  will report on this effect in an upcoming publication.}
\bibitem[{ref({\natexlab{g}})}]{ref:logznote}
\bibinfo{note}{We can eliminate the scale factor 
of $\log Z$ for any unweighted graph. However, here we 
wish to demonstrate the full weighted scaling, so
almost all of our unweighted graph examples include 
the $\log Z$ scale factor.}
\bibitem[{ref({\natexlab{h}})}]{ref:nmicalcnote}
\bibinfo{note}{For the $I_N$, $V$, and $I$ calculations, 
we are generally comparing configurations that are 
somewhat similar. The confusion matrix can have at most 
$N$ non-zero entries; therefore, we use a pseudo-sparse 
matrix representation to calculate these measures 
usually in $O(N)$. Worst case (pathological) behavior 
for initialization is $O(Nq)$. The calculation cost
can be as fast as $O(q)$ for strongly correlated systems. 
The total cost for Step~\ref{step:correlations} of the 
algorithm will typically scale as $O(Nr^2\log N)$ where 
the user has control over the number of replicas $r$
and $\log N$ is the estimated scaling for the required 
number of resolutions.}
\bibitem[{ref({\natexlab{m}})}]{ref:entropydifferent}
\bibinfo{note}{As a specific realization, 
consider a symmetric adjacency matrix $A$
for a four node graph given by 
$A_{12}=A_{13}=A_{14}=A_{23}=A_{24}=1$.
The permutation $P_{34}$ that exchanges $3$ with $4$
leaves $A$ invariant,
%whereas $A$ is not invariant under the permutation $P_{14}$. 
%If $A$ is invariant under a permutation of $i$ with $j$ 
%(e.g, $3$ with $4$), 
so the graph itself is invariant.
If we have two solutions $B$ and $C$ that 
place $i$ and $j$ in different communities then 
$I_N(B,C)\neq 1$ despite the invariance 
of $A$ under the permutation.
Unless one can distinguish between the nodes via 
external means, community groupings of the four nodes such 
as $(123)(4)$ and $(124)(3)$ correspond 
to the same breaking of the lattice
(the graph is symmetric under the permutation),
but they  have a relative mutual information that 
differs from the self mutual information 
[$(8-3 \log 3)/4$ vs $(10-6 \log 3)/4$ 
for the two groupings respectively].}
\end{thebibliography}

\end{document}